\documentclass[aps,twocolumn,floatfix,superscriptaddress,longbibliography]{revtex4-2}
\usepackage[utf8]{inputenc}
\usepackage{times}
\usepackage{tikz}
\usepackage[T1]{fontenc}

\usepackage{bm}
\usepackage{times}
 \usepackage{algorithm,algorithmic}

\usepackage{mathtools}
\usepackage{amsmath}
\usepackage[shortlabels]{enumitem}

\usepackage{graphicx,epic,eepic,epsfig,amsmath,latexsym,amssymb,verbatim,color}

\usepackage{bbm}

\usepackage{float}
\usepackage{tikz}
\usetikzlibrary{chains}
\usetikzlibrary{fit}
\usepackage{pgflibraryarrows}		
\usepackage{pgflibrarysnakes}		

\usepackage{epsfig}
\usetikzlibrary{shapes.symbols,patterns} 
\usepackage{pgfplots}

\usepackage[strict]{changepage}
\usepackage{hyperref}
\hypersetup{colorlinks=true,citecolor=blue,linkcolor=blue,filecolor=blue,urlcolor=blue,breaklinks=true}

\usepackage[marginal]{footmisc}
\usepackage{url}
\usepackage{theorem}
 
\def\squareforqed{\hbox{\rlap{$\sqcap$}$\sqcup$}}
\def\qed{\ifmmode\squareforqed\else{\unskip\nobreak\hfil
\penalty50\hskip1em\null\nobreak\hfil\squareforqed
\parfillskip=0pt\finalhyphendemerits=0\endgraf}\fi}
\def\endenv{\ifmmode\;\else{\unskip\nobreak\hfil
\penalty50\hskip1em\null\nobreak\hfil\;
\parfillskip=0pt\finalhyphendemerits=0\endgraf}\fi}

\newcounter{remark}

\newcounter{example}

\mathchardef\ordinarycolon\mathcode`\:
\mathcode`\:=\string"8000
\def\vcentcolon{\mathrel{\mathop\ordinarycolon}}
\begingroup \catcode`\:=\active
  \lowercase{\endgroup
  \let :\vcentcolon
  }

\usepackage{cleveref}
\usepackage{graphicx}
\usepackage{xcolor}

\RequirePackage[framemethod=default]{mdframed}
\newmdenv[skipabove=7pt,
skipbelow=7pt,
backgroundcolor=darkblue!15,
innerleftmargin=5pt,
innerrightmargin=5pt,
innertopmargin=5pt,
leftmargin=0cm,
rightmargin=0cm,
innerbottommargin=5pt,
linewidth=1pt]{tBox}

\newmdenv[skipabove=7pt,
skipbelow=7pt,
backgroundcolor=blue2!25,
innerleftmargin=5pt,
innerrightmargin=5pt,
innertopmargin=5pt,
leftmargin=0cm,
rightmargin=0cm,
innerbottommargin=5pt,
linewidth=1pt]{dBox}
\newmdenv[skipabove=7pt,
skipbelow=7pt,
backgroundcolor=darkkblue!15,
innerleftmargin=5pt,
innerrightmargin=5pt,
innertopmargin=5pt,
leftmargin=0cm,
rightmargin=0cm,
innerbottommargin=5pt,
linewidth=1pt]{sBox}
\definecolor{darkblue}{RGB}{0,76,156}
\definecolor{darkkblue}{RGB}{0,0,153}
\definecolor{blue2}{RGB}{102,178,255}
\definecolor{darkred}{RGB}{195,0,0}

\newcommand{\nc}{\newcommand}
\nc{\rnc}{\renewcommand}
\nc{\beg}{\begin{equation}}
\nc{\eeq}{{\end{equation}}}
\nc{\beqa}{\begin{eqnarray}}
\nc{\eeqa}{\end{eqnarray}}
\nc{\lbar}[1]{\overline{#1}}
\nc{\bra}[1]{\langle#1|}
\nc{\ket}[1]{|#1\rangle}
\nc{\ketbra}[2]{|#1\rangle\!\langle#2|}
\nc{\braket}[2]{\langle#1|#2\rangle}

\nc{\proj}[1]{| #1\rangle\!\langle #1 |}
\nc{\avg}[1]{\langle#1\rangle}
\nc{\rank}{\operatorname{Rank}}
\nc{\smfrac}[2]{\mbox{$\frac{#1}{#2}$}}
\nc{\tr}{\operatorname{Tr}}
\nc{\ox}{\otimes}
\nc{\dg}{\dagger}
\nc{\dn}{\downarrow}
\nc{\cA}{{\cal A}}
\nc{\cB}{{\cal B}}
\nc{\cC}{{\cal C}}
\nc{\cD}{{\cal D}}
\nc{\cE}{{\cal E}}
\nc{\cF}{{\cal F}}
\nc{\cG}{{\cal G}}
\nc{\cH}{{\cal H}}
\nc{\cI}{{\cal I}}
\nc{\cJ}{{\cal J}}
\nc{\cK}{{\cal K}}
\nc{\cL}{{\cal L}}
\nc{\cM}{{\cal M}}
\nc{\cN}{{\cal N}}
\nc{\cO}{{\cal O}}
\nc{\cP}{{\cal P}}
\nc{\cQ}{{\cal Q}}
\nc{\cR}{{\cal R}}
\nc{\cS}{{\cal S}}
\nc{\cT}{{\cal T}}
\nc{\cV}{{\cal V}}
\nc{\cX}{{\cal X}}
\nc{\cY}{{\cal Y}}
\nc{\cZ}{{\cal Z}}
\nc{\cW}{{\cal W}}
\nc{\csupp}{{\operatorname{csupp}}}
\nc{\qsupp}{{\operatorname{qsupp}}}
\nc{\var}{{\operatorname{var}}}
\nc{\rar}{\rightarrow}
\nc{\lrar}{\longrightarrow}
\nc{\polylog}{{\operatorname{polylog}}}
\nc{\wt}{{\operatorname{wt}}}
\nc{\av}[1]{{\left\langle {#1} \right\rangle}}
\nc{\supp}{{\operatorname{supp}}}

\nc{\argmin}{{\operatorname{argmin}}}

\def\x{\xi}

\nc{\RR}{{{\mathbb R}}}
\nc{\CC}{{{\mathbb C}}}
\nc{\FF}{{{\mathbb F}}}
\nc{\NN}{{{\mathbb N}}}
\nc{\ZZ}{{{\mathbb Z}}}
\nc{\PP}{{{\mathbb P}}}
\nc{\QQ}{{{\mathbb Q}}}
\nc{\UU}{{{\mathbb U}}}
\nc{\EE}{{{\mathbb E}}}
\nc{\id}{{\operatorname{id}}}

\nc{\CHSH}{{\operatorname{CHSH}}}

\nc{\be}{\begin{equation}}
\nc{\ee}{{\end{equation}}}
\nc{\bea}{\begin{eqnarray}}
\nc{\eea}{\end{eqnarray}}
\nc{\<}{\langle}
\rnc{\>}{\rangle}
\nc{\rU}{\mbox{U}}

\nc{\ob}[1]{#1}

\nc{\SEP}{{\text{\rm SEP}}}
\nc{\NS}{{\text{\rm NS}}}
\nc{\LOCC}{{\text{\rm LOCC}}}
\nc{\PPT}{{\text{\rm PPT}}}
\nc{\EXT}{{\text{\rm EXT}}}
\nc{\Sym}{{\operatorname{Sym}}}

\nc{\ERLO}{{E_{\text{r,LO}}}}
\nc{\ERLOCC}{{E_{\text{r,LOCC}}}}
\nc{\ERPPT}{{E_{\text{r,PPT}}}}
\nc{\ERLOCCinfty}{{E^{\infty}_{\text{r,LOCC}}}}
\nc{\Aram}{{\operatorname{\sf A}}}

\usepackage{tikz}
\usepackage{hyperref}
\hypersetup{colorlinks=true,citecolor=blue,linkcolor=blue,filecolor=blue,urlcolor=blue,breaklinks=true}

\makeatletter
\def\grd@save@target#1{%
  \def\grd@target{#1}}
\def\grd@save@start#1{%
  \def\grd@start{#1}}
\tikzset{
  grid with coordinates/.style={
    to path={%
      \pgfextra{%
        \edef\grd@@target{(\tikztotarget)}%
        \tikz@scan@one@point\grd@save@target\grd@@target\relax
        \edef\grd@@start{(\tikztostart)}%
        \tikz@scan@one@point\grd@save@start\grd@@start\relax
        \draw[minor help lines,magenta] (\tikztostart) grid (\tikztotarget);
        \draw[major help lines] (\tikztostart) grid (\tikztotarget);
        \grd@start
        \pgfmathsetmacro{\grd@xa}{\the\pgf@x/1cm}
        \pgfmathsetmacro{\grd@ya}{\the\pgf@y/1cm}
        \grd@target
        \pgfmathsetmacro{\grd@xb}{\the\pgf@x/1cm}
        \pgfmathsetmacro{\grd@yb}{\the\pgf@y/1cm}
        \pgfmathsetmacro{\grd@xc}{\grd@xa + \pgfkeysvalueof{/tikz/grid with coordinates/major step}}
        \pgfmathsetmacro{\grd@yc}{\grd@ya + \pgfkeysvalueof{/tikz/grid with coordinates/major step}}
        \foreach \x in {\grd@xa,\grd@xc,...,\grd@xb}
        \node[anchor=north] at (\x,\grd@ya) {\pgfmathprintnumber{\x}};
        \foreach \y in {\grd@ya,\grd@yc,...,\grd@yb}
        \node[anchor=east] at (\grd@xa,\y) {\pgfmathprintnumber{\y}};
      }
    }
  },
  minor help lines/.style={
    help lines,
    step=\pgfkeysvalueof{/tikz/grid with coordinates/minor step}
  },
  major help lines/.style={
    help lines,
    line width=\pgfkeysvalueof{/tikz/grid with coordinates/major line width},
    step=\pgfkeysvalueof{/tikz/grid with coordinates/major step}
  },
  grid with coordinates/.cd,
  minor step/.initial=.2,
  major step/.initial=1,
  major line width/.initial=2pt,
}
\makeatother

\usepackage{thmtools}
\usepackage{thm-restate}
\usepackage{etoolbox}
\makeatletter
\def\problem@s{}
\newcounter{problems@cnt}

\newcommand{\allproblems}{\problem@s}
\makeatother

\definecolor{beamer}{rgb}{0.2,0.2,0.7}
\definecolor{colorone}{rgb}{1,0.36,0.03}
\definecolor{colortwo}{rgb}{0.4,0.77,0.17}
\definecolor{colorthree}{rgb}{0.01,0.51,0.93}
\definecolor{colorfour}{rgb}{0.47,0.26,0.58}
\definecolor{colorfive}{rgb}{0.12,0.55,0.16}
\usepackage{tcolorbox}
\usepackage{relsize}
\usepackage{graphicx}
\usepackage{subfigure}
\usepackage{booktabs}
\usepackage{array}
\newcommand{\angs}{$\mathring{A}$}
\newcommand{\invangs}{$\mathring{A}^{-1}$}
\allowdisplaybreaks

\begin{document}

\title{N-representable one-electron reduced density matrix reconstruction with frozen core electrons}
 
\author{Sizhuo Yu}
\email{sizhuo.yu@centralesupelec.fr}
\affiliation{Université Paris-Saclay, CentraleSupélec, CNRS, Laboratoire SPMS, F 91190 Gif-sur-Yvette, France}

\author{Jean-Michel Gillet}
\affiliation{Université Paris-Saclay, CentraleSupélec, CNRS, Laboratoire SPMS, F 91190 Gif-sur-Yvette, France}

\begin{abstract}
Recent advances in quantum crystallography have shown that, beyond conventional charge density refinement, a one-electron reduced density matrix (1-RDM) satisfying N-representability conditions can be reconstructed using jointly experimental X-ray structure factors (XSF) and directional Compton profiles (DCP) through semi-definite programming.
So far, such reconstruction methods for  1-RDM, not constrained to idempotency, had been tested only on a toy model system (CO$_2$).
In this work, a new method is assessed on crystalline urea (CO(NH$_2$)$_2$) using static (0 K) and dynamic (50 K) artificial-experimental data.
An improved model, including symmetry constraints and frozen-core electron contribution, is introduced to better handle the increasing system complexity. 
Reconstructed 1-RDMs, deformation densities and DCP anisotropy are analyzed, and it is demonstrated that the changes in the model significantly improve the reconstruction's quality against insufficient information and data corruption.
The robustness of the model and the strategy are thus shown to be well-adapted to address the reconstruction problem from actual experimental scattering data.
\end{abstract}

\date{\today}

\maketitle


\section{Introduction}\label{sec-intro}
While N-electron wave-functions provide the most complete and exact description of electronic structure in crystals, their experimental determination is still out of reach due to their exponentially large complexity for real systems. 
Moreover, in Coulson's words: "a conventional many-electron wave-function tells us more than we need to know."\cite{coulsonPresentStateMolecular1960} 
It is then worth considering the one(two)-electron reduced density matrices (1,2-RDM) as compact substitutes for wave-functions since they involve significantly fewer parameters. As of today, the incompleteness of N-representability conditions\cite{Liu2007}, which ensure that a reduced density matrix can be associated with a complete N-body density matrix, and the lack of experimental observables with sufficient information content still pose daunting obstacles to the reconstruction of 2-RDMs.
Therefore, 1-RDMs, which do not suffer from the same impediment and still contain valuable quantum mechanical information, are considered suitable candidates for modelling electron behaviour from experimental data. 
The reconstruction process, however, remains a challenging task. 
Firstly, N-representability conditions still need to be fulfilled for an experimentally reconstructed 1-RDM to be physically meaningful. Secondly, from a pure measurement perspective, as the 1-RDM contains both position and momentum space information, it cannot be obtained using a single experimental technique to this day and to the best of our knowledge.

The challenge of 1-RDM reconstruction from experimental data was initiated by Clinton and coworkers in the 1960s using a drastic idempotency condition as a means to ensure N-representability 
 \cite{Clinton1969_1, Clinton1969_2, Clinton1969_3, Clinton1969_4, Clinton1969_5}.
Based on a series of works combining position and momentum space data on isolated atoms, Schmider and coworkers \cite{schmiderReconstructionOneParticle1992} argued that the idempotency condition would hinder the recovery of electron (static and dynamical) correlation effect in the reconstructed density matrix. 
The potential presence of such information in position space was recently confirmed by an X-ray constrained wave-function refinement on urea and alanine \cite{hupfEffectsExperimentallyObtained2023}. 
The authors argue that evidence of significant deviation from the Hartree-Fock description can be found using high-resolution X-ray diffraction structure factors. Any single-determinant-based model would forbid access to such subtle features in the data. 
Adopting a formal perspective, Mazziotti and coworkers discussed \cite{mazziotti2007reduced} different strategies to include N-representability conditions in a series of articles and proposed a semidefinite programming (SDP) formulation of the  1,2-RDM reconstruction problem \cite{foleyMeasurementdrivenReconstructionManyparticle2012}. On more practical grounds,
following Schmider and coworkers' seminal work,  several papers reported the joint use of X-ray diffraction structure factors (SF) and directional Compton profiles (DCP) to explore different non-single-determinant models and strategies for 1-RDM modelling in both magnetic and non-magnetic molecular compounds \cite{schmiderAtomicOrbitalsCompton1993,schmiderInferenceOneParticleDensity1993,schwarzDensityMatricesPosition1994b,schmiderLowMomentumElectrons1996,gueddidaDevelopmentJointRefinement2018,gueddidaJointRefinementModel2018,debruyneInferringOneelectronReduced2020,launayNRepresentableOneelectronReduced2021}. However, all SDP-based reconstruction attempts of 1-RDM, which have been put forward, were applied to isolated atoms or molecules with at most 2 or 3 atoms.

The present work further investigates the 1-RDM reconstruction problem in molecular crystals by building upon the convex optimization approach put forward in \cite{debruyneInferringOneelectronReduced2020,launayNRepresentableOneelectronReduced2021}, scaling up the system size from modest dry ice (CO$_2$) to the more realistic urea (CO(NH$_2$)$_2$) crystal. The purpose is thus to demonstrate the potential of an improved method more suitable to practical applications and its aptness to compensate for sparse momentum space data. To address the challenges posed by a significant increase in system size, we propose the implementation of symmetry constraints and the possibility of freezing core-electron contributions. For the first time, approximate energy and virial ratio are used to determine the optimal data set for the 1-RDM model refinement.

This article is structured as follows: In Sec. 2, we explain how the 1-RDM reconstruction can be formulated as a convex optimization problem, with the N-representability condition, symmetry and frozen core electrons as convex constraints. The method used for reconstruction, deconvoluted from thermal motion, is briefly reviewed. In Sec. 3, we showcase the importance of the joint use of position and momentum space data even when Compton scattering data is suspected to be poorly informative. Additional degradation due to noise and temperature effects and the improved robustness using symmetry and frozen core constraints are illustrated. The conclusion and future directions are given in the last section.

\section{Methods}
\subsection{1-RDM reconstruction using least-square fitting}

For a spin-traced(spin-free) pure-state N-electron system, the 1-RDM can be derived by integrating out the $N-1$ coordinates of the N-electron density matrix, i.e.
\begin{equation}
    \Gamma^{(1)}(\mathbf{r}, \mathbf{r}') =  N \int \psi(\mathbf{r}, \mathbf{r_2}, ... \mathbf{r_N} )\psi^* (\mathbf{r}', \mathbf{r}_2, ..., \mathbf{r_N}) d\mathbf{r}_2 ... d\mathbf{r}_N.
    \label{eq_pure1rdm}
\end{equation}
where $\psi(\mathbf{r}, \mathbf{r_2}, ... \mathbf{r_N} )$ is the pure-state N-electron wavefunction. A mixed-state system 1-RDM is a mere convex combination of pure-state 1-RDMs. 

It is well-known \cite{lowdinQuantumTheoryManyParticle1955a} that the 1-RDM can be conveniently approximated using a discrete one-electron basis set $\{\phi_i\}$ as
\begin{equation}
    \Gamma^{(1)}(\mathbf{r}, \mathbf{r}') = \sum_{ij} P_{ij} \phi_i(\mathbf{r}) \phi^*_j (\mathbf{r}').
    \label{eq_1rdmP}
\end{equation}
If the basis set is kept fixed, the 1-RDM is determined once the population matrix $\mathbf{P}$ in \eqref{eq_1rdmP} is found. The number of parameters in the model is thus solely conditioned by the size of the population matrix and, therefore, by the number of basis functions. In this work, the basis functions are atomic orbitals, but plane waves could also be considered, if needed, for strongly delocalised electron systems.

The 1-RDM is directly connected to the  mean electron density distribution in position space through its  diagonal elements 
\begin{equation}\label{1RDM2density}
 \rho(\mathbf{r}) = \Gamma^{(1)}(\mathbf{r}, \mathbf{r}).  
\end{equation}
Furthermore, the 1-RDM encapsulates momentum space information through a 6D Fourier-Dirac transform \cite{Weyrich1996} 
\begin{equation}
    n(\mathbf{p}) = \frac{1 }{ (2\pi\hbar)^3}\int\Gamma^{(1)}(\mathbf{r}, \mathbf{r}+\mathbf{t})e^{-i\mathbf{p}\cdot\mathbf{t}/\hbar} d^3t d^3r,
\end{equation} 
with $n(\mathbf{p})$ being the momentum density.
This double connection to both axes of phase space strongly suggests there is little hope of reconstructing a good quality 1-RDM from data provided by a single experimental technique.

Thanks to very efficient refinement methods and models \cite{gattiModernChargeDensityAnalysis2012a}, high-resolution X-ray structure factors (SF), which are obtained by elastic coherent X-ray diffraction, are almost routinely used in the reconstruction of position space electron density. Using \eqref{1RDM2density}, the relationship between SF and 1-RDM is simply
\begin{equation}
    F(\mathbf{q}) = \int \Gamma^{(1)}(\mathbf{r}, \mathbf{r}) \exp (-i \mathbf{r} \cdot \mathbf{q}) d \mathbf{r}
    \label{eq_sfraw}.
\end{equation}
On the other hand, directional Compton profiles (DCP) are measured by deep inelastic incoherent X-ray scattering. Within the Impulse Approximation \cite{phillipsXRayDeterminationElectron1968}, they give access to projections of momentum space electron density, i.e.
\begin{align}
 J(q,\mathbf{u})& = \int n(\mathbf{p}) \delta(\mathbf{p} \cdot \mathbf{u} - q) d \mathbf{p}\\ &=\frac{1 }{ 2\pi\hbar}\int\Gamma^{(1)}(\mathbf{r}, \mathbf{r}+t \mathbf{u})e^{-iqt/\hbar} dt d^3r,
    \label{eq_dcpraw}
\end{align}
where $\mathbf{u}$ is the unit vector giving the direction in momentum space onto which the electron density is projected. It is colinear with the scattering vector of the Compton measurement.

The model used in this work is based on expression \eqref{eq_1rdmP}. The determination of the best population matrix given a set of SF and DCP thus requires expressing experimental observable values as functions of matrix $\mathbf{P}$ using the operator form
\begin{equation}
F(\mathbf{q}) = \text{Tr} (\mathbf{F}_\mathbf{q} \mathbf{P})~~\text{and}~~
J(\mathbf{q}) = \text{Tr} (\mathbf{J}_\mathbf{q} \mathbf{P}),
\label{eq_FJops}
\end{equation}
with $\mathbf{F}_\mathbf{q}$ and $\mathbf{J}_\mathbf{q}$ being the SF and DCP operators respectively. For conciseness, $\mathbf{q}$ stands for $(q,\mathbf{u})$ in the Compton profile matrix element of \eqref{eq_FJops}. To proceed any further, the operators $\mathbf{F}_\mathbf{q}$ and $ \mathbf{J}_\mathbf{q}$  matrix elements need to be written in the basis-set representation as
\begin{align}
\begin{split}
(\mathbf{F}_\mathbf{q})_{ij} = \int \phi_i^*(\mathbf{r})\phi_j(\mathbf{r})e^{-i\mathbf{q}\cdot\mathbf{r}} d^3r,\\
(\mathbf{J}_\mathbf{q})_{ij} = \frac{1}{2\pi \hbar}\int \phi_i^*(\mathbf{r})\phi_j(\mathbf{r}+t \mathbf{u})e^{-iqt} dtd^3r.
\end{split}
\label{eq_FJelt}
\end{align}

In this work, particular attention has been paid to the reliability of the final 1-RDM reconstruction. If one assumes that error bars on data points are uncorrelated and follow a normal distribution law, for an unbiased model, the most probable population matrix $\mathbf{P}$ is found by solving the minimization problem.
\begin{equation}
    \text{argmin}_{\mathbf{P}} ~ \sum_i \left (\frac{ \text{Tr}(\mathbf{O}_i \mathbf{P}) - O^{\text{exp}}_i }{\sigma_i}\right )^2,
    \label{eq_chi2}
\end{equation}
where the model is expected to yield the mean value for each observable datum represented by $\mathbf{O}_i$ while its actual experimental measurement gives $ O^{\text{exp}}_i$ with the associated estimated variance $\sigma_i^2$. In our case, each data point originates either from X-ray diffraction or Compton scattering measurements so that $\mathbf{O}_i = \mathbf{F}_{\mathbf{q}_i}$ or, $\mathbf{O}_j =\mathbf{J}_{\mathbf{q}_j}$ for different scattering vectors. 
In the present work, for a given basis set of Gaussian contracted Slater-type orbitals, the closed form of each matrix element \eqref{eq_FJelt} is calculated prior to refinement using a Mathematica code \cite{Mathematica}. 

The minimization of \eqref{eq_chi2} is a convex least-squares fitting problem. The following section will explain how, together with the necessary N-representability conditions, the reconstruction problem falls into a convex optimization problem called semidefinite programming \cite{boyd2004convex}.

\subsection{Constraints: N-representability, symmetry and frozen core}
\label{sec_constraint} 
The N-representability conditions must be satisfied to ensure that the population matrix yields a physically meaningful density matrix. It is worth noting that the N-representability conditions are significantly more difficult if one requires the system to be in a pure state instead of a statistical mixture of quantum states.  \textit{Pure} N-representability and \textit{ensemble} N-representability are generally employed to distinguish the respective situations \cite{Chakraborty2015Nrep}. We have chosen to consider the latter case for both practical reasons and because the system cannot always be exactly in its ground state, without interacting with the environment. Consequently, for an \textit{ensemble} N-representable 1-RDM, the population matrix $\mathbf{P}^\perp$ for a closed-shell system, associated with an orthonormal basis set, must satisfy the following constraints,
\begin{subequations}\label{Nrepcond10}
    \begin{align}
      \mathbf{P}^\perp &\succcurlyeq 0,\label{Nrepcond11} \\
        2 \mathbf{I} - \mathbf{P}^\perp &\succcurlyeq 0, \label{Nrepcond12}\\
        \text{Tr}(\mathbf{P}^\perp) &= N,
    \label{Nrepcond13}
    \end{align}
\end{subequations}
together with the obvious condition that
$\mathbf{P}^\perp $ 
being Hermitian. Here $\mathbf{I}$ is the identity matrix, and the symbol $\succcurlyeq$ means the matrix is semidefinite positive, which is equivalent to stating that all eigenvalues are non-negative. Hence, constraint (11b) requires the eigenvalues of $\mathbf{P}^\perp$ to be smaller than 2. As previously mentioned, the present basis set is made of Slater-type atomic orbitals (expressed as Gaussian contractions), which are not mutually orthogonal. A Lowdin orthogonalization is thus performed on the atomic-orbital basis set prior to the reconstruction.

All constraints listed in \eqref{Nrepcond10} are convex; thus, the convexity of the minimization of Eq.~\eqref{eq_chi2} is preserved. Moreover, the semi-definite positivity of $\mathbf{P}^\perp$ imposed in \eqref{Nrepcond11} makes it possible to use the tools of semi-definite programming \cite{foleyMeasurementdrivenReconstructionManyparticle2012, debruyneInferringOneelectronReduced2020}. Access to the solution is thereby significantly facilitated.  

The model developed in this work is specifically adapted to molecular crystals for which a single group of atoms can be considered to form a specific entity. It is assumed that this group, referred to as "the molecule", does not share any charge with other entities in the same or neighbouring unit cells. The 1-RDM model is thus a mere molecular 1-RDM onto which translation and rotational symmetry operations can be applied to generate the density matrix of the entire crystal. These operations are fully taken into account in the present work.

The symmetry invariance at the molecular level can also be considered. The population matrix is thus required to be a direct sum of matrices in the invariant subspaces of each symmetry operator. In other words, $\mathbf{P}$ should be block-diagonal when using the symmetry-adapted orbitals as the new basis, i.e. $\mathbf{S}^{T} \mathbf{P} \mathbf{S} = \bigoplus_{j=1}^{n} \mathbf{P}_j$ where $\mathbf{S}$ transforms the basis of atomic orbitals into symmetry adapted orbitals, $n$ the number of irreducible representations and $\mathbf{P}_j$ the block matrices associated with each irreducible representation.

The new model also allows for freezing core-electron contributions. It effectively reduces the model's active space, hence the number of parameters to be determined in the population matrix.
As a consequence, illustrated in the next section, the computational cost is lowered, and the robustness of the result is improved against noise contamination and thermal-induced effects.
It can be best observed on core electrons' spatial density distribution, contributing to sharp peaks near each nucleus. 
Therefore, accurately reproducing such features for a population matrix model would require knowledge of high-$q$ structure factors, which may present an experimental challenge at usual temperatures. 
Here, an alternate but common approach was chosen.  A single-determinant calculation of the wave-function is performed from which core-electron molecular orbitals are extracted to construct an approximate core-electron density matrix. As a result, the model population matrix is given by $\mathbf{P}^\perp = \mathbf{P}'^\perp + \mathbf{P}_{\text{core}}^\perp$, with $\mathbf{P}_{\text{core}}^\perp$ being the frozen-core-electron population matrix. 
The latter represents a fixed number of electrons and is -by construction- idempotent. The optimization is thus forced to search for the optimal solution in the subspace orthogonal to that spanned by the core electron's orbitals if the N-representability on the total 1-RDM is to be preserved.

Combining symmetry and frozen-core conditions and assuming a non-magnetic system, the N-representability constraints become
\begin{subequations}\label{eq_cons_symcore_all}
    \begin{align}
    \mathbf{P}'^\perp &\succcurlyeq 0, 
    \label{eq_cons_symcore_1}
    \\
    2 \mathbf{I} - \mathbf{P}'^\perp - \mathbf{P}^\perp_{\text{core}}&\succcurlyeq 0, \\
    \text{Tr}(\mathbf{P}'^\perp + \mathbf{P}^\perp_{\text{core}}) &= N \\
    \mathbf{S'}^{T} (\mathbf{P}'^\perp + \mathbf{P}^\perp_{\text{core}}) \mathbf{S'} &= \bigoplus_{j=1}^{n} \mathbf{P}_j,
    \label{eq_cons_symcore_4}
    \end{align}
\end{subequations}
with $\mathbf{P}'^\perp$ being the population matrix for valence electrons and $N$ the total electron number for a single molecule. $\mathbf{S'}$ transforms the orthogonalized atomic basis into the symmetry adapted basis. 
We remark that with \eqref{eq_cons_symcore_1} - \eqref{eq_cons_symcore_4}, the optimization problem can still be modelled with SDP.
In this work, the constrained optimization problem is solved from the closed form of our model using the CVXPY package \cite{diamond2016cvxpy}.

\begin{figure}[!htbp]
    \centering
 \includegraphics[width=9cm]{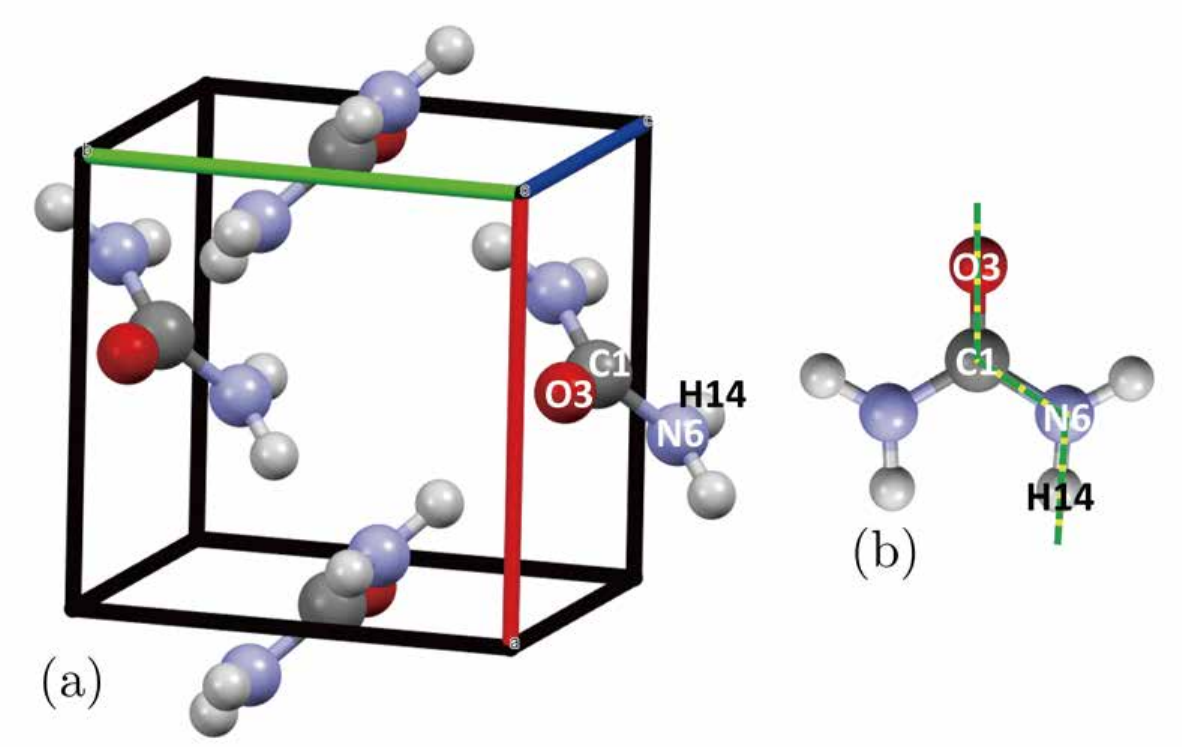}
    \caption{a) The unit cell of urea crystal ($P\overline{4}21m$ with a=5.66 \angs, c=4.71 \angs) b) In dashed-green the path along which the 1-RDM values displayed in this article have been computed. The path is a succession of segments passing through O-C-N-H atoms.}
    \label{fig1}
\end{figure}

\subsection{Reconstruction from non-zero temperature data}
\label{sec_dw}
It must be noted that the reconstruction method is inherently temperature-independent, since the 1-RDM describes both mixed states and pure states. However, comparison with first-principle calculations is generally best done at the zero-Kelvin limit. It is thus helpful to deconvolute thermal motion effects to recover the ideal static 1-RDM. In this work, it is assumed that, given the large photon-electron energy transfer involved in the Compton scattering process,
DCPs are hardly affected by nuclear agitation at reasonably low temperature \cite{sternemannInfluenceLatticeDynamics2000,dugdaleThermalDisorderCorrelation1998,matsudaRayComptonScattering2020}. Therefore, temperature-induced alteration of experimental data is only taken into consideration for X-ray structure factors. In this case, the model is modified so that the SF matrix elements include an anisotropic Debye-Waller factor.
\begin{equation}\label{DWSFmatrix}
    (\mathbf{F}_\mathbf{q})_{ij}= e^{-\mathbf{q}\cdot\widehat {B}_a\cdot\mathbf{q}}\int \phi_i^*(\mathbf{r})\phi_j(\mathbf{r})e^{-i\mathbf{q}\cdot\mathbf{r}} d^3r
\end{equation}
where $\widehat {B}_a$ is the thermal displacement tensor for nucleus $a$ on which both basis functions $\phi_i$ and $\phi_j$ are centred. No change is applied when the basis functions are associated with different atoms. More sophisticated temperature schemes are worth considering \cite{stevensCalculationDynamicElectron1977}. For example, the Mulliken partitioning approach to two-centre contribution was implemented in our previous work \cite{launayNRepresentableOneelectronReduced2021} and should be used with real data. However, the usual independent-atom model was chosen to prevent unfair similarity with the computational method used to generate the reference data \cite{erbaAccurateDynamicalStructure2013a}. It has been checked that this simple approach allows for a fair deconvolution of thermal agitation effects when data is not contaminated with noise.

\section{Results}

The model explained above is well suited to molecular crystals and should be assessed for realistic systems. In particular, for such an approach which combines different experiments, it is necessary to evaluate the impact of data quality on the 1-RDM reconstruction.

The urea crystal (CO(NH$_2$)$_2$) has been chosen for two specific reasons: firstly, it has long been considered a "standard" test system in the field of charge density reconstruction. Several bond types are represented, among which highly mobile and delocalized electron density contributes to non-linear optical properties \cite{cassidyNonlinearOpticalProperties1979,westComprehensiveAnalysisTerms2015}. Secondly, because of the interest it has attracted over the years, high-quality structure factors \cite{zavodnikElectronDensityStudy1999,birkedalChargeDensityUrea2004} and Compton profiles data \cite{shuklaHydrogenBondingUrea2001} are available from the literature. It thus positions urea as a legitimate candidate for a first phase-space-derived reconstruction of experimental 1-RDM on a molecular compound. Additionally, the urea molecule is significantly larger than our previous test systems and possibly one of the largest molecules onto which Compton measurement has ever been reported \cite{shuklaHydrogenBondingUrea2001}. It can thus be considered a significant step in the quest for 1-RDM reconstruction.
This paper is the last stage of model calibration before a final reconstruction from true experimental data is undertaken.

We use here the same strategy for model assessment as for smaller systems, and described in previous papers \cite{debruyneInferringOneelectronReduced2020,launayNRepresentableOneelectronReduced2021}: a reference 1-RDM is obtained from a periodic DFT calculation using the B3LYP functional \cite{beckeDensityFunctionalThermochemistry1993} and a pob-DZVP basis set \cite{peintingerConsistentGaussianBasis2013} using the CRYSTAL14 program \cite{Dovesi2014Crystal}. The nuclei positions are those given by \cite{Worsham1986} and derived from neutron diffraction data.
Artificial-experimental data points are then generated based on this DFT-derived 1-RDM. 50-K-structure-factors are computed up to $\sin\theta/\lambda=$1.1 \invangs after atomic displacement parameters have been obtained using the dedicated option of CRYSTAL14 \cite{erbaAccurateDynamicalStructure2013a}.
Compton profiles are little affected by thermal motion at such low temperatures, and no particular treatment is applied in their case.
The CRYSTAL14 SF and DCP values are considered ideal mean values on which a Gaussian noise distribution is centred for each data point.
Consequently, noise-contaminated data is also considered in our test reconstructions.
The reconstructed density matrix is obtained by determining a population matrix for a basis of poorer quality than that employed for artificial-data generation. An inevitable bias in the model is therefore introduced.
The basis set for the 1-RDM model is thus taken as a simple 6-31G basis set, with additional $p$-orbitals on hydrogen atoms.

\subsection{Reconstructions from ideal data}
\begin{figure}[!tp]
    \centering
    \includegraphics[width=0.5\textwidth]{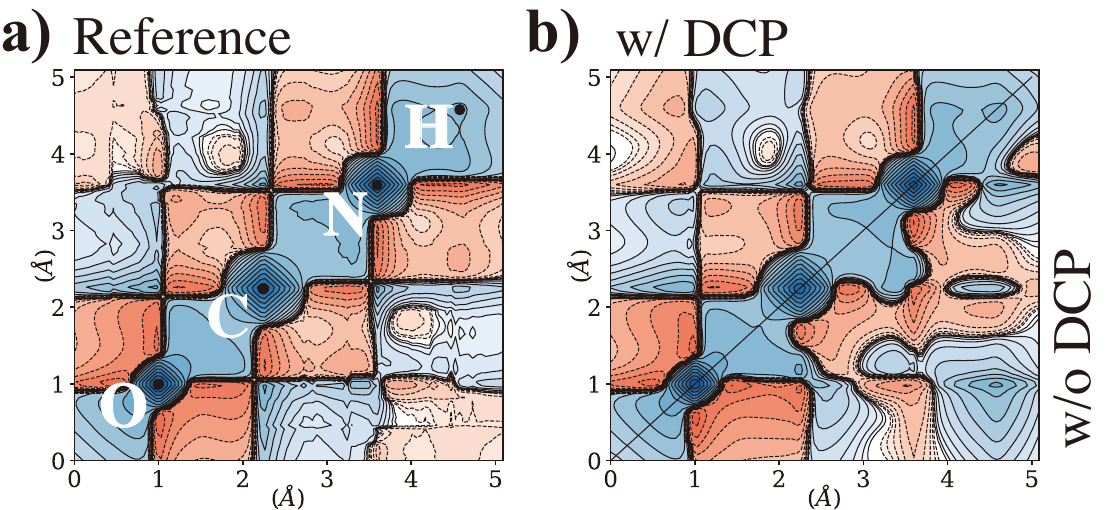}
    \caption{a) The reference 1-RDM along the O-C-N-H path was calculated from CRYSTAL14. The reconstructed 1-RDMs with and without the DCP artificial-data are shown respectively in the upper left and lower right corner of (b). The contours are drawn at $\pm 10^{-2} \times 2^n e \cdot \text{\angs}^{-3}, ~n 
    \in [0, 20]$ where the positive(negative) contours are shown in solid (dashed) lines with blue (red) shades. }
    \label{fig2}
\end{figure}

The best reconstruction result is expected when data is obtained without thermal motion and noise. 
The use of artificial data cannot be circumvented to test this optimal case. 
Observing what type of reconstruction results from the sole use of X-ray diffraction data is then quite illustrative. 
The artificial-experimental set includes 3627 SF with $\sin\theta/\lambda < 1.1 $ \invangs. 
Inspection of the 1-RDM $\Gamma(\mathbf{r}, \mathbf{r}')$ on the O-C-N-H path as a 2D function clearly shows that the SF-only derived 1-RDM lacks most of the off-diagonal regions important features (Fig. \ref{fig2}). It is consistent with conclusions drawn from previous works on a much smaller system. 
In such a case, inferring the off-diagonal region is inherently difficult because SF are solely related to the position space density, therefore to the diagonal component $\rho(\mathbf{r}) = \Gamma(\mathbf{r}, \mathbf{r})$. Only constraints on the model are likely to improve the off-diagonal description. This is an important criterion to assess the quality of the 1-RDM reconstruction since, in essence, off-diagonal parts are conditioned by the bonding mechanisms and how different locations interfere to shape the wave-function.

A second step is to include noise-free Compton data in the observables. In all the following cases, 8 non-equivalent crystalline directions are used ([100], [110], [111], [210], [211], [310], [311], [321]). For each direction, data points are taken every 0.1 a.u.. This value corresponds to usual Compton spectrometer resolutions and prevents significant correlation between consecutive points. The maximum momentum value is set to 10 a.u.. The data set thus contains 800 DCP values in total. Obviously, a noise-free refinement case does not justify any weighting scheme, and the $\sigma_i$ in the objective function \eqref{eq_chi2} are uniformly taken to be 1.

As displayed in Fig.~\ref{fig2}(b), the reconstructed 1-RDM now exhibits very marginal deviation from the reference. Slight differences persist in the off-diagonal regions $\Gamma(r, r'\neq r)$. 
A discrepancy is observed when the reconstructed 1-RDM are visualized along the two different O-C-N-H paths. Such a discrepancy is corrected once the symmetry restriction is imposed.

The virial ratio $-V / 2T$ is calculated for the reconstructed 1-RDMs, where the two-electron potential energy is estimated using the 2-RDM expression ansatz $\Gamma^{(2)}(\mathbf{r'}_1,\mathbf{r'}_2;\mathbf{r}_1,\mathbf{r}_2)=\Gamma^{(1)}(\mathbf{r}_1,\mathbf{r'}_1)\Gamma^{(1)}(\mathbf{r'}_2,\mathbf{r}_2)-\Gamma^{(1)}(\mathbf{r'}_1,\mathbf{r}_2)\Gamma^{(1)}(\mathbf{r'}_2,\mathbf{r}_1)$. 
The virial ratios for reconstruction with and without DCP are 0.996 and 0.934, respectively, confirming the role of Compton data in reaching a more pertinent solution.
The distinction between the two reconstructions showcases the importance of momentum space measurement even for a system like urea crystal, where the DCP anisotropy does not exceed 1\%  of the total electron number (see Fig.~\ref{fig_dcp}). 
Note that the good post-refinement virial ratio is a mere consequence of the reconstruction quality and did not require any ad-hoc constraint in our model or the objective function.

In the following paragraphs, possible sources of reconstruction errors will be discussed in more detail, and emphasis will be put on techniques for improving the model's robustness.

\subsection{Closer to real life: noise and temperature effects}\label{section_noisetemp}

\begin{figure}[!htb]
    \centering
    \includegraphics[width=0.5\textwidth]{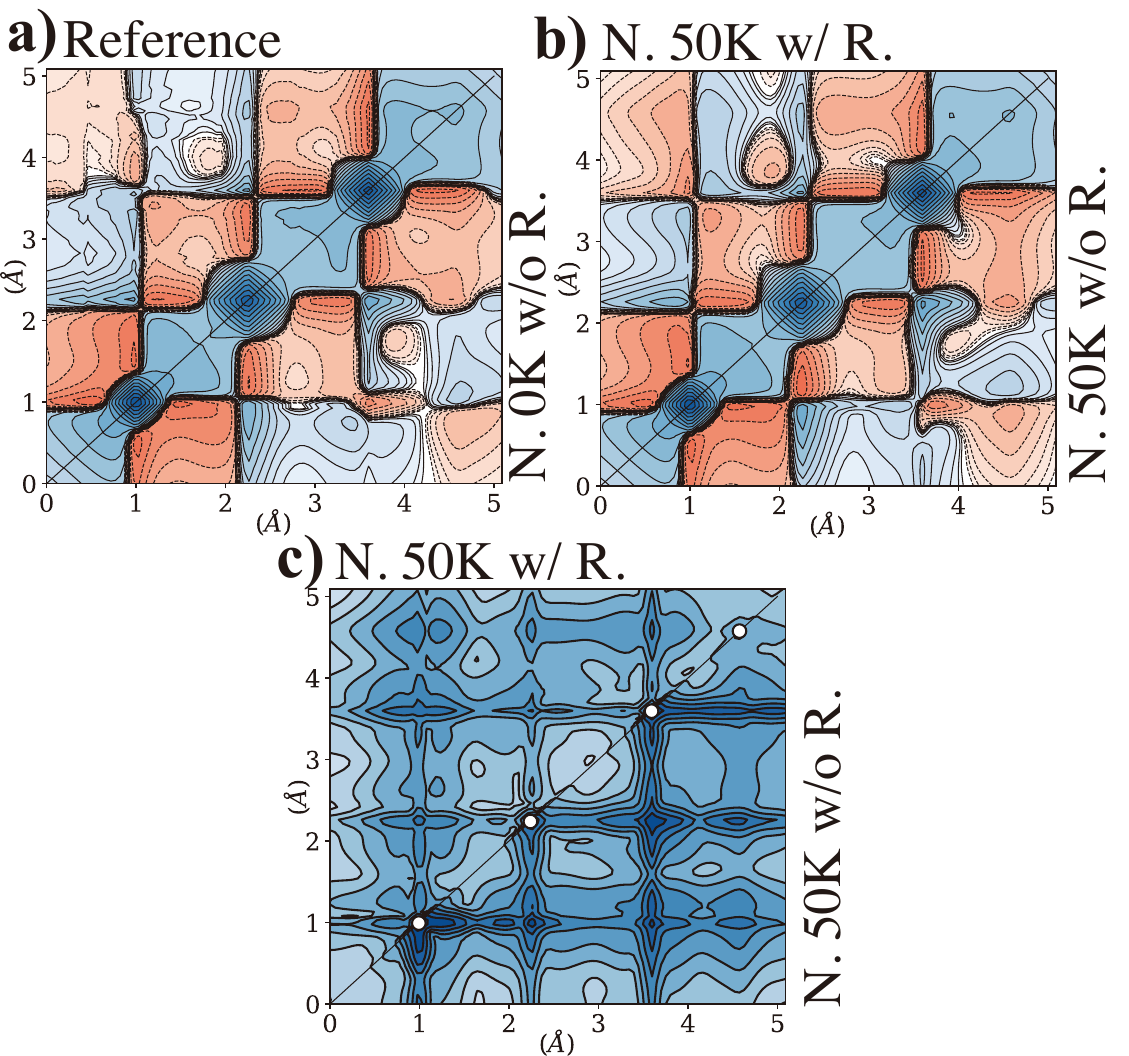}
    \caption{a) The reference (upper left) and reconstructed (lower right) 1-RDM $\Gamma(\mathbf{r}, \mathbf{r}')$ with 0 K 1\% noisy data.  b) The reconstruction 1-RDMs from 50 K 1\% noisy data with (upper left) and without (lower right) restrictions. c) Estimated standard deviations for reconstructions shown in (b).
    \\
    The contours for (a) and (b) are the same as in Fig.~\ref{fig2},  and the positive(negative) contours are shown in solid (dashed) lines with blue (red) shades. For (c) the contours are drawn at $10^{-4} \times 2^n e \cdot \text{\angs}^{-3}, ~n \in [0, 12].$
    \\
    (N. 0K w/o R. = Noised 0 K without Restriction)}
    \label{fig3}
\end{figure}

When real experimental data is used, noise contamination cannot be avoided. This section first considers the effect of statistical noise and, as a common practice, assumes no bias in the model. Then, the thermal motion of nuclei is introduced, and we study how it combines with statistical noise to deteriorate the reconstructed 1-RDM further. 

Artificial data is now contaminated by a random noise generated according to a Gaussian law. 
For example, the SF data values become $F'(\mathbf{q}) = F(\mathbf{q}) + n \times \epsilon (\mathbf{q})$ with $~\epsilon \sim \mathcal{N}(0, \vert F (\mathbf{q}) \vert )$. The noise level is chosen to be of the order of 1\% by setting $n=0.01$. 
A similar procedure is applied to the DCP values. 
Notice that, given the weak Compton anisotropy in this system, the chosen noise level wipes out most of the directional information from the Compton scattering spectrum. We have found that such a noise model results in highly unbalanced weight in the objective function. Hence, an unweighted version of \eqref{eq_chi2} is used in practice.

\begin{figure}
    \centering
    \includegraphics[width=0.5\textwidth]{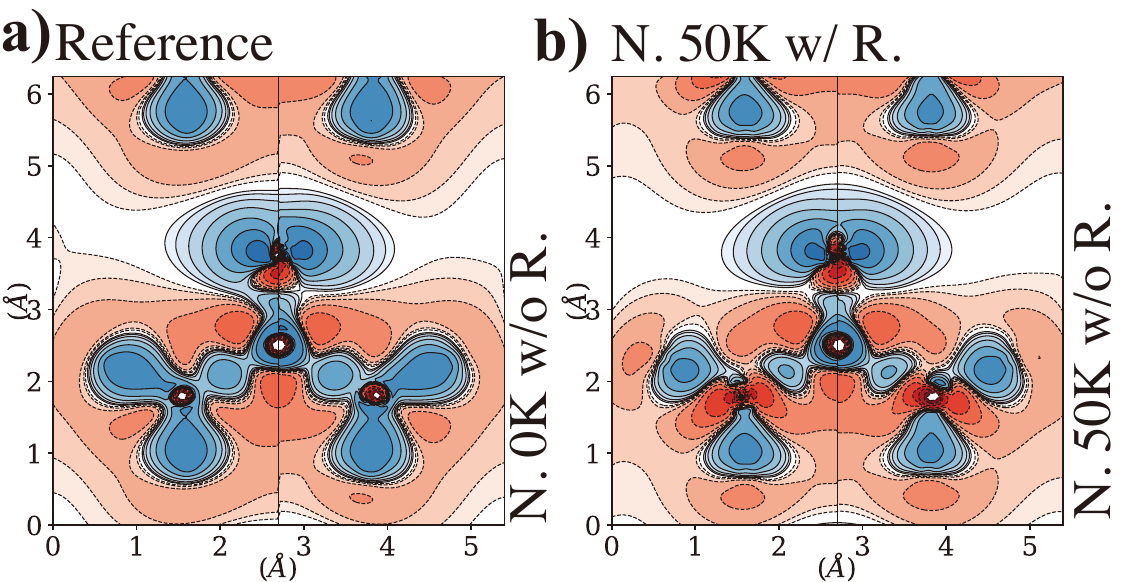}
    \caption{a) Deformation density on C-O-N reference plan (left) and reconstruction with 0 K 1\% noisy data (right). b) Deformation density reconstructed from 50 K 1\% noisy data with (left) and without (right) core electron and symmetry restrictions. \\
    The contours are drawn at the same levels as Fig.~\ref{fig2}.}
    \label{fig_deform}
\end{figure}

\begin{figure}
    \centering
    \includegraphics[width=0.5\textwidth]{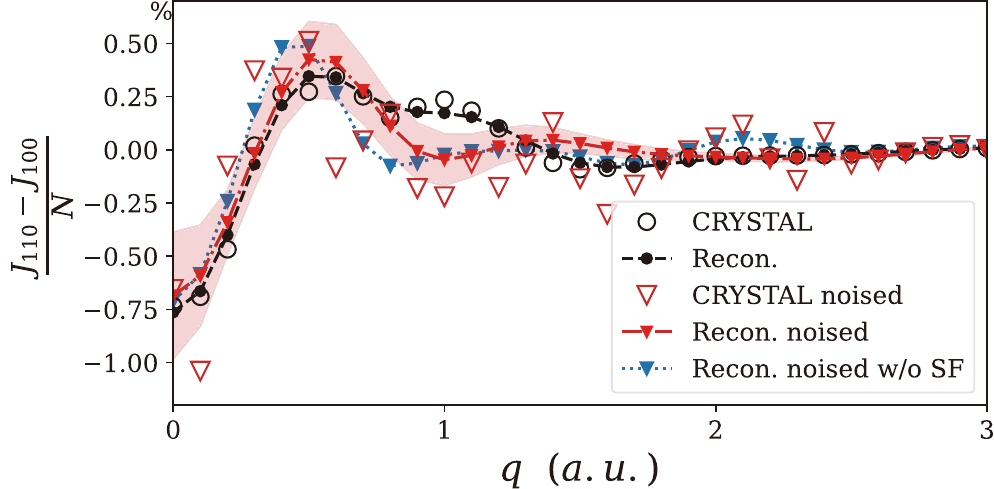}
    \caption{Reference (dots) and reconstructed (lines) DCP anisotropy at [110] direction without (circles) and with 1\% noise (triangles). The dotted line shows the reconstruction without the use of SF artificial-data. The purple shaded area indicates the standard deviation of reconstruction upon resampling from the noise distribution. }
    \label{fig_dcp}
\end{figure}

As expected, the 1-RDM reconstruction from noisy data now exhibits stronger deviations from the reference. 
It can be seen in Fig. \ref{fig3} (a). The modest discrepancies in the diagonal part of the RDM can be emphasized by looking at the electron deformation density displayed in Fig.~ \ref{fig_deform} (a). As a reminder, the deformation density is the difference between the total electron density and the sum of independent atoms densities, i.e. promolecular density. The latter is obtained from CRYSTAL14 software using the same basis set as the reference calculation (pob-DZVP).
As anticipated, discrepancies are more significant in the off-diagonal region of the 1-RDM (Fig.~\ref{fig3} (a)). The model, constrained by the N-representability conditions, obviously struggles to get sensible information from the weak Compton anisotropies buried under the noise. It is evidenced by the noticeable mismatch of anisotropy oscillations shown in  Fig.~\ref{fig_dcp} (filled red triangles). Nevertheless, it can be seen that the deviation of reconstructed DCP anisotropy is still moderate, possibly due to the information carried by structure factors data. This assumption is validated after observing further deterioration when SF are removed from the data (filled blue triangles).

For the proposed model, deconvolution of temperature effects constitutes a difficult challenge.
In contrast to most common electron density reconstructions, using, for example, the widely spread kappa-refinement pseudo-atom multipolar model \cite{hansenTestingAsphericalAtom1978, gattiModernChargeDensityAnalysis2012a}, our approach to 1-RDM determination relies on a linear expression \eqref{eq_1rdmP} which, combined with linear constraints (Sec. \ref{sec_constraint}), makes it possible to use positive semi-definite programming methodology.
Insertion of the Debye-Waller formulation to account for the thermal effect destroys such a linearity. While an alternative formulation is currently under development, it was decided for the present work to explore the possibility of treating sequentially both problems. First, an ab-initio 1-RDM is computed with the basis set of the model.
Then, the Atomic Displacement Parameters (ADP) are determined from high-order structure factors ($\sin\theta/\lambda>0.7$ \invangs) as explained in Sec.~\ref{sec_dw}. Then, these ADP values $\mathbf{B}$ are fixed and incorporated into the model. Therefore, the model remains linear for the 1-RDM refinement step, since a mere factor $e^{-\mathbf{q} \cdot \mathbf{B} \cdot \mathbf{q}}$  is added to the SF operators.
The quality of such ADPs depends heavily on the model basis-set, and one cannot expect the refined $\mathbf{P}$ matrix to be exempt from thermal motion contamination.
As shown in the lower panel of Fig.~\ref{fig3}(b) and Fig.~\ref{fig_deform}(b), the reconstructed 1-RDM and deformation density continue to worsen, which is clear evidence that the thermal motion effect has not been thoroughly deconvoluted.
Although Compton data is assumed to be unperturbed at such low temperatures, the off-diagonal region continues deteriorating. It must be attributed to the sparsity of reliable information from momentum space, which cannot be compensated for by the SFs. 
Our independent-atom Debye-Waller description's poor performance is clearly shown by unphysical electron depletion in the vicinity of nitrogen centres shown in Fig.~\ref{fig_deform}(b). Further, it is confirmed by the significant differences between the refined and reference (from CRYSTAL14) ADP for the nitrogen nuclei (about 25\% discrepancy). When real data is involved, this very crude scheme will necessitate the addition of the previously mentioned two-centre terms in \eqref{DWSFmatrix} and a more thorough inclusion of the Debye-Waller contribution in the general refinement.  The feature is currently being implemented.

\subsection{Further restrictions: frozen-core and symmetry}

In the previous section, we discussed how the combination of noise and thermal motion affects the 1-RDM reconstruction using \eqref{eq_1rdmP}. 
To mitigate such problems, a possible method is to reduce the degree of freedom of the model, thus making it more robust against noise contamination.
As introduced in Sec.~\ref{sec_constraint}, one would naturally first invoke the necessity of applying symmetry restrictions to the model.
An overall improvement in reconstruction quality is observed as unnecessary free parameters are eliminated.

A further limitation of the active space is obtained by freezing the core electron contribution to the density matrix. 
This well-spread procedure does not affect our ability to absorb momentum space data, which primarily describes delocalised valence electrons. 
On the SF side, freezing the core component of the 1-RDM helps stabilise the refinement against high-order reflections, which are the most affected by the noise and nuclear motion, while preserving nearly all the model's flexibility. 
In this subsection, we report the impact of such a scheme under the non-ideal reconstruction scenarios.

When the frozen-core and symmetry restrictions are added to those concerning N-representability, Fig.~\ref{fig3} (b - upper panel) shows that the distortion in the reconstructed 1-RDM is greatly reduced.
In this case, even in the presence of noise and thermal agitation, the model catches most of the features observed in the reference 1-RDM (Fig.~\ref{fig3}(a). Note that the most significant discrepancy is in the off-diagonal region corresponding to the long-range interaction between hydrogen and carbon, which are second neighbours. Such a striking improvement confirms that limiting the active space can effectively improve the reconstruction's robustness against noise. 
The standard deviation on the reconstructed 1-RDMs with and without additional constraints (Fig. \ref{fig3}(c)) was estimated upon resampling from the Gaussian noise distribution. 
It is observed that the restriction of active space distinctly diminishes the uncertainty of the reconstruction.

Interestingly, such a betterment in the 1-RDM modelling brings only minor changes to the resulting deformation density near the nuclei. Similarly, no major improvement is observed for the DCP anisotropy reconstruction (see Supplementary Information). This seemingly paradoxical observation can be resolved when adopting an optimization problem perspective.

As mentioned earlier, the 1-RDM reconstruction was defined through \eqref{eq_chi2} as a least-squares minimization problem given the SF and DCP data.
Therefore, introducing constraints such as \eqref{eq_cons_symcore_1} - \eqref{eq_cons_symcore_4} can only result in a new optimal solution with higher $\chi^2$ value, i.e. a worse fit to the SF and DCP.
Consequently, the DCP anisotropies and deformation density are not likely to be improved because they only depend on our ability to fit the Compton data and a set of Fourier coefficients of the electron density.
However, a 1-RDM is a function in 6-D space which contains more information than its limited number of projections given by the data values.
In such a case, it is well-founded to believe that restricting the size of solution space effectively regularizes the model, giving it stronger predictive ability.

\begin{figure}[!htb]
    \centering
    \includegraphics[width=0.46\textwidth]{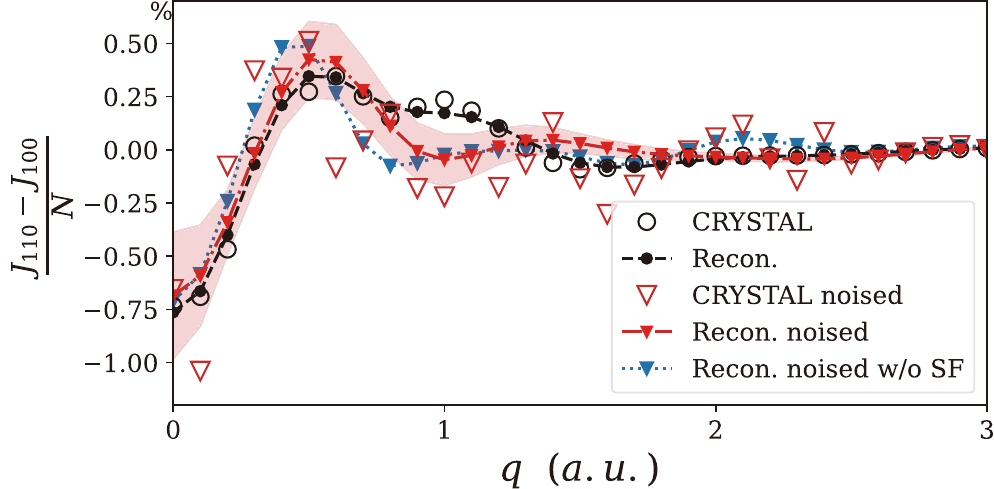}
    \caption{Mean field energy (see text) of one urea molecule reconstructed from 1\% noisy data. Dashed and solid lines with circle and triangle data points represent the reconstruction with 0 K and 50 K SF data, respectively, and with identical Compton data. Blue lines show the reconstruction with no additional constraints. Light blue and red lines show the results when symmetry and core electron constraints are used. The purple dotted line shows the virial ratio of a 0 K 1\% noisy data reconstruction with additional restrictions.}
    \label{fig5}
\end{figure}

Estimating the total electron energy from experimental SF is a well-known, difficult challenge.
Adding Compton scattering information does not significantly facilitate the task. However, on a mere relative scale, the energy criterion can be employed to compare the performances of different refinement strategies.
Throughout this work, a recurrent question has been to evaluate the optimal cut-off value in $\sin\theta/\lambda$ for the structure factors. In a perfect world, free from thermal motion, high-Miller-indices reflections should be retained as long as they rise above statistical noise.
The solid curve in Fig. \ref{fig5}  shows that, for an ideal 0 K set of SF, the total electron energy stabilizes for any cut-off value above 0.7 \invangs.
It is no longer true when data values are affected by temperature agitation.
In the 50 K case (dashed curve), SF corresponding to $\sin\theta/\lambda > 0.7$ \invangs contribute to a significant deterioration of the reconstruction from an electron-energy perspective.
As shown in Fig. \ref{fig5} (all dashed-curves), minimum energy is reached when only reflections lower than 0.7 \invangs are included in the set. Then, as one increases the Ewald sphere radius,  the total energy starts rising continuously. This confirms that the additional high-order reflections, which are the most affected by thermal motion, are not sufficiently well deconvoluted by the one-centre Debye-Waller model and merely contribute to perturbing the refinement process.

When symmetry enforcement alone is applied, an overall betterment can already be observed under 0 K and 50 K scenarios. 
More remarkably, introducing an additional frozen-core component not only further reduces the perturbation instilled by high-order reflections but also dramatically improves the reconstruction ability when only a small amount of reflection data is available.
Thus, both restrictions effectively increase the stability of the model by filtering out most of the perturbation brought by thermal motion and noise contamination and by reducing the number of unnecessary free parameters.
In addition, the behaviour of the fully restricted model in the reduced $q_{\text{max}}$ domain suggests the possibility of reconstructing the 1-RDM with a limited amount of low-angle SF data, those that describe the most diffused electrons.

Let us insist that the Hartree-Fock-like energy computed here is only meaningful as an indicator of the reconstruction quality since it uses both position and momentum space electron densities. However, due to the intrinsic difficulty of predicting energy from 1-RDMs, the question of whether one could accurately determine the total (or interaction) energy from scattering experiments should be left for more careful examination and discussion.

\section{Conclusion and Discussion}
In this work, an improved 1-RDM reconstruction method has been tested on a system which is significantly larger than those previously investigated \cite{debruyneInferringOneelectronReduced2020, launayNRepresentableOneelectronReduced2021}. 
The crucial role played by momentum space information, originating from Compton scattering data, is confirmed. It is instrumental in the quality of the reconstruction, even when the weak anisotropy is buried under statistical noise.
The two main additions to the model, symmetry restrictions and frozen-core contributions, are evidenced to drastically stabilize the 1-RDM reconstruction process against statistical noise and temperature effects.
Meanwhile, with no additional constraints, it is shown that the resulting energies, evaluated from the modelled 1-RDM, closely satisfy the virial theorem. 
As a consequence, the approximated total energy and virial ratio were found to be valuable indicators to identify an optimal portion of the Ewald sphere, which balances pertinent information and noise contamination.

However, proper deconvolution of temperature-induced nuclear motion remains a challenging problem. 
In the current approach, two main obstacles have been identified. Firstly, our choice of limiting the flexibility of the temperature model and the basis set to avoid bias in the assessment. 
Secondly, the necessity of keeping the 1-RDM model linear.  
Both inevitably led to strong discrepancies in atomic displacement parameters but allowed for a reliable assessment of the model's stability. Moreover, we have good reasons to believe that using a non-linear version of the optimization, including two-centre temperature factors and a better basis set, will drastically improve the performance when real experimental data is considered.

The current method for 1-RDM reconstruction is essentially a statistical inference procedure. Therefore, the quality of its outcome depends not only on the data distribution but also on the prior distribution of the model. In the present stage, a uniform prior was used, which means no prior knowledge is assumed. In the future, one could consider a more informed prior, for example, a Gaussian distribution centred on a lower-level theory calculation. The use of additional priors will help the model's performance, especially in the case of poor data quality.

Finally, our results illustrate that 1-RDM reconstruction is achievable for a system of moderate size from X-ray structure factors and directional Compton profile measurements, even when the momentum space information is drastically limited. In the next step, such a method can be readily applied to actual experimental data.

\textbf{Acknowledgements: } The authors thank Devinder Sivia, Pietro Cortona and Julie McDonald for their insightful comments and discussions. S. Y. gratefully acknowledges funding from the Chinese Scholarship Council. Part of the computations are conducted on the clusters of Paris-Saclay University, which is gratefully acknowledged.

\bibliographystyle{apsrev4-1}
\bibliography{references}

\begin{thebibliography}{42}%
\makeatletter
\providecommand \@ifxundefined [1]{%
 \@ifx{#1\undefined}
}%
\providecommand \@ifnum [1]{%
 \ifnum #1\expandafter \@firstoftwo
 \else \expandafter \@secondoftwo
 \fi
}%
\providecommand \@ifx [1]{%
 \ifx #1\expandafter \@firstoftwo
 \else \expandafter \@secondoftwo
 \fi
}%
\providecommand \natexlab [1]{#1}%
\providecommand \enquote  [1]{``#1''}%
\providecommand \bibnamefont  [1]{#1}%
\providecommand \bibfnamefont [1]{#1}%
\providecommand \citenamefont [1]{#1}%
\providecommand \href@noop [0]{\@secondoftwo}%
\providecommand \href [0]{\begingroup \@sanitize@url \@href}%
\providecommand \@href[1]{\@@startlink{#1}\@@href}%
\providecommand \@@href[1]{\endgroup#1\@@endlink}%
\providecommand \@sanitize@url [0]{\catcode `\\12\catcode `\$12\catcode
  `\&12\catcode `\#12\catcode `\^12\catcode `\_12\catcode `\%12\relax}%
\providecommand \@@startlink[1]{}%
\providecommand \@@endlink[0]{}%
\providecommand \url  [0]{\begingroup\@sanitize@url \@url }%
\providecommand \@url [1]{\endgroup\@href {#1}{\urlprefix }}%
\providecommand \urlprefix  [0]{URL }%
\providecommand \Eprint [0]{\href }%
\providecommand \doibase [0]{http://dx.doi.org/}%
\providecommand \selectlanguage [0]{\@gobble}%
\providecommand \bibinfo  [0]{\@secondoftwo}%
\providecommand \bibfield  [0]{\@secondoftwo}%
\providecommand \translation [1]{[#1]}%
\providecommand \BibitemOpen [0]{}%
\providecommand \bibitemStop [0]{}%
\providecommand \bibitemNoStop [0]{.\EOS\space}%
\providecommand \EOS [0]{\spacefactor3000\relax}%
\providecommand \BibitemShut  [1]{\csname bibitem#1\endcsname}%
\let\auto@bib@innerbib\@empty
\bibitem [{\citenamefont {Coulson}(1960)}]{coulsonPresentStateMolecular1960}%
  \BibitemOpen
  \bibfield  {author} {\bibinfo {author} {\bibfnamefont {C.~A.}\ \bibnamefont
  {Coulson}},\ }\href@noop {} {\bibfield  {journal} {\bibinfo  {journal} {Rev.
  Mod. Phys.}\ }\textbf {\bibinfo {volume} {32}},\ \bibinfo {pages} {170}
  (\bibinfo {year} {1960})}\BibitemShut {NoStop}%
\bibitem [{\citenamefont {Liu}\ \emph {et~al.}(2007)\citenamefont {Liu},
  \citenamefont {Christandl},\ and\ \citenamefont {Verstraete}}]{Liu2007}%
  \BibitemOpen
  \bibfield  {author} {\bibinfo {author} {\bibfnamefont {Y.~K.}\ \bibnamefont
  {Liu}}, \bibinfo {author} {\bibfnamefont {M.}~\bibnamefont {Christandl}}, \
  and\ \bibinfo {author} {\bibfnamefont {F.}~\bibnamefont {Verstraete}},\
  }\href {\doibase 10.1103/PhysRevLett.98.110503} {\bibfield  {journal}
  {\bibinfo  {journal} {Phys. Rev. Lett.}\ }\textbf {\bibinfo {volume} {98}},\
  \bibinfo {pages} {110503} (\bibinfo {year} {2007})}\BibitemShut {NoStop}%
\bibitem [{\citenamefont {Clinton}\ \emph
  {et~al.}(1969{\natexlab{a}})\citenamefont {Clinton}, \citenamefont
  {Nakhleh},\ and\ \citenamefont {Wunderlich}}]{Clinton1969_1}%
  \BibitemOpen
  \bibfield  {author} {\bibinfo {author} {\bibfnamefont {W.~L.}\ \bibnamefont
  {Clinton}}, \bibinfo {author} {\bibfnamefont {J.}~\bibnamefont {Nakhleh}}, \
  and\ \bibinfo {author} {\bibfnamefont {F.}~\bibnamefont {Wunderlich}},\
  }\href {\doibase 10.1103/PhysRev.177.1} {\bibfield  {journal} {\bibinfo
  {journal} {Phys. Rev.}\ }\textbf {\bibinfo {volume} {177}},\ \bibinfo {pages}
  {1} (\bibinfo {year} {1969}{\natexlab{a}})}\BibitemShut {NoStop}%
\bibitem [{\citenamefont {Clinton}\ \emph
  {et~al.}(1969{\natexlab{b}})\citenamefont {Clinton}, \citenamefont {Galli},\
  and\ \citenamefont {Massa}}]{Clinton1969_2}%
  \BibitemOpen
  \bibfield  {author} {\bibinfo {author} {\bibfnamefont {W.~L.}\ \bibnamefont
  {Clinton}}, \bibinfo {author} {\bibfnamefont {A.~J.}\ \bibnamefont {Galli}},
  \ and\ \bibinfo {author} {\bibfnamefont {L.~J.}\ \bibnamefont {Massa}},\
  }\href {\doibase 10.1103/PHYSREV.177.7} {\bibfield  {journal} {\bibinfo
  {journal} {Phys. Rev.}\ }\textbf {\bibinfo {volume} {177}},\ \bibinfo {pages}
  {7} (\bibinfo {year} {1969}{\natexlab{b}})}\BibitemShut {NoStop}%
\bibitem [{\citenamefont {Clinton}\ \emph
  {et~al.}(1969{\natexlab{c}})\citenamefont {Clinton}, \citenamefont
  {Henderson},\ and\ \citenamefont {Prestia}}]{Clinton1969_3}%
  \BibitemOpen
  \bibfield  {author} {\bibinfo {author} {\bibfnamefont {W.~L.}\ \bibnamefont
  {Clinton}}, \bibinfo {author} {\bibfnamefont {G.~A.}\ \bibnamefont
  {Henderson}}, \ and\ \bibinfo {author} {\bibfnamefont {J.~V.}\ \bibnamefont
  {Prestia}},\ }\href {\doibase 10.1103/PhysRev.177.13} {\bibfield  {journal}
  {\bibinfo  {journal} {Phys. Rev.}\ }\textbf {\bibinfo {volume} {177}},\
  \bibinfo {pages} {13} (\bibinfo {year} {1969}{\natexlab{c}})}\BibitemShut
  {NoStop}%
\bibitem [{\citenamefont {Clinton}\ and\ \citenamefont
  {Lamers}(1969)}]{Clinton1969_4}%
  \BibitemOpen
  \bibfield  {author} {\bibinfo {author} {\bibfnamefont {W.~L.}\ \bibnamefont
  {Clinton}}\ and\ \bibinfo {author} {\bibfnamefont {G.~B.}\ \bibnamefont
  {Lamers}},\ }\href {\doibase 10.1103/PhysRev.177.19} {\bibfield  {journal}
  {\bibinfo  {journal} {Phys. Rev.}\ }\textbf {\bibinfo {volume} {177}},\
  \bibinfo {pages} {19} (\bibinfo {year} {1969})}\BibitemShut {NoStop}%
\bibitem [{\citenamefont {Clinton}\ \emph
  {et~al.}(1969{\natexlab{d}})\citenamefont {Clinton}, \citenamefont {Galli},
  \citenamefont {Henderson}, \citenamefont {Lamers}, \citenamefont {Massa},\
  and\ \citenamefont {Zarur}}]{Clinton1969_5}%
  \BibitemOpen
  \bibfield  {author} {\bibinfo {author} {\bibfnamefont {W.~L.}\ \bibnamefont
  {Clinton}}, \bibinfo {author} {\bibfnamefont {A.~J.}\ \bibnamefont {Galli}},
  \bibinfo {author} {\bibfnamefont {G.~A.}\ \bibnamefont {Henderson}}, \bibinfo
  {author} {\bibfnamefont {G.~B.}\ \bibnamefont {Lamers}}, \bibinfo {author}
  {\bibfnamefont {L.~J.}\ \bibnamefont {Massa}}, \ and\ \bibinfo {author}
  {\bibfnamefont {J.}~\bibnamefont {Zarur}},\ }\href {\doibase
  10.1103/PHYSREV.177.27} {\bibfield  {journal} {\bibinfo  {journal} {Phys.
  Rev.}\ }\textbf {\bibinfo {volume} {177}},\ \bibinfo {pages} {27} (\bibinfo
  {year} {1969}{\natexlab{d}})}\BibitemShut {NoStop}%
\bibitem [{\citenamefont {Schmider}\ \emph {et~al.}(1992)\citenamefont
  {Schmider}, \citenamefont {Smith},\ and\ \citenamefont
  {Weyrich}}]{schmiderReconstructionOneParticle1992}%
  \BibitemOpen
  \bibfield  {author} {\bibinfo {author} {\bibfnamefont {H.}~\bibnamefont
  {Schmider}}, \bibinfo {author} {\bibfnamefont {J.}~\bibnamefont {Smith},
  \bibfnamefont {Vedene~H.}}, \ and\ \bibinfo {author} {\bibfnamefont
  {W.}~\bibnamefont {Weyrich}},\ }\href {\doibase 10.1063/1.462256} {\bibfield
  {journal} {\bibinfo  {journal} {J. Chem. Phys.}\ }\textbf {\bibinfo {volume}
  {96}},\ \bibinfo {pages} {8986} (\bibinfo {year} {1992})}\BibitemShut
  {NoStop}%
\bibitem [{\citenamefont {Hupf}\ \emph {et~al.}(2023)\citenamefont {Hupf},
  \citenamefont {Kleemiss}, \citenamefont {Borrmann}, \citenamefont {Pal},
  \citenamefont {Krzeszczakowska}, \citenamefont {Woińska}, \citenamefont
  {Jayatilaka}, \citenamefont {Genoni},\ and\ \citenamefont
  {Grabowsky}}]{hupfEffectsExperimentallyObtained2023}%
  \BibitemOpen
  \bibfield  {author} {\bibinfo {author} {\bibfnamefont {E.}~\bibnamefont
  {Hupf}}, \bibinfo {author} {\bibfnamefont {F.}~\bibnamefont {Kleemiss}},
  \bibinfo {author} {\bibfnamefont {T.}~\bibnamefont {Borrmann}}, \bibinfo
  {author} {\bibfnamefont {R.}~\bibnamefont {Pal}}, \bibinfo {author}
  {\bibfnamefont {J.~M.}\ \bibnamefont {Krzeszczakowska}}, \bibinfo {author}
  {\bibfnamefont {M.}~\bibnamefont {Woińska}}, \bibinfo {author}
  {\bibfnamefont {D.}~\bibnamefont {Jayatilaka}}, \bibinfo {author}
  {\bibfnamefont {A.}~\bibnamefont {Genoni}}, \ and\ \bibinfo {author}
  {\bibfnamefont {S.}~\bibnamefont {Grabowsky}},\ }\href {\doibase
  10.1063/5.0138312} {\bibfield  {journal} {\bibinfo  {journal} {J. Chem.
  Phys}\ }\textbf {\bibinfo {volume} {158}},\ \bibinfo {pages} {124103}
  (\bibinfo {year} {2023})}\BibitemShut {NoStop}%
\bibitem [{\citenamefont {Mazziotti}(2007)}]{mazziotti2007reduced}%
  \BibitemOpen
  \bibfield  {author} {\bibinfo {author} {\bibfnamefont {D.~A.}\ \bibnamefont
  {Mazziotti}},\ }\href@noop {} {\emph {\bibinfo {title}
  {Reduced-density-matrix mechanics: with applications to many-electron atoms
  and molecules}}},\ Vol.\ \bibinfo {volume} {134}\ (\bibinfo  {publisher}
  {Wiley Online Library},\ \bibinfo {year} {2007})\BibitemShut {NoStop}%
\bibitem [{\citenamefont {Foley}\ and\ \citenamefont
  {Mazziotti}(2012)}]{foleyMeasurementdrivenReconstructionManyparticle2012}%
  \BibitemOpen
  \bibfield  {author} {\bibinfo {author} {\bibfnamefont {J.~J.}\ \bibnamefont
  {Foley}}\ and\ \bibinfo {author} {\bibfnamefont {D.~A.}\ \bibnamefont
  {Mazziotti}},\ }\href {\doibase 10.1103/PhysRevA.86.012512} {\bibfield
  {journal} {\bibinfo  {journal} {Phys. Rev. A}\ }\textbf {\bibinfo {volume}
  {86}},\ \bibinfo {pages} {012512} (\bibinfo {year} {2012})}\BibitemShut
  {NoStop}%
\bibitem [{\citenamefont {Schmider}\ and\ \citenamefont
  {Smith}(1993)}]{schmiderAtomicOrbitalsCompton1993}%
  \BibitemOpen
  \bibfield  {author} {\bibinfo {author} {\bibfnamefont {H.}~\bibnamefont
  {Schmider}}\ and\ \bibinfo {author} {\bibfnamefont {V.~H.}\ \bibnamefont
  {Smith}, \bibfnamefont {Jr.}},\ }\href {\doibase 10.1515/zna-1993-1-242}
  {\bibfield  {journal} {\bibinfo  {journal} {Zeitschrift für Naturforschung
  A}\ }\textbf {\bibinfo {volume} {48}},\ \bibinfo {pages} {221} (\bibinfo
  {year} {1993})}\BibitemShut {NoStop}%
\bibitem [{\citenamefont {Schmider}\ \emph {et~al.}(1993)\citenamefont
  {Schmider}, \citenamefont {Smith},\ and\ \citenamefont
  {Weyrich}}]{schmiderInferenceOneParticleDensity1993}%
  \BibitemOpen
  \bibfield  {author} {\bibinfo {author} {\bibfnamefont {H.}~\bibnamefont
  {Schmider}}, \bibinfo {author} {\bibfnamefont {V.~H.}\ \bibnamefont {Smith},
  \bibfnamefont {Jr.}}, \ and\ \bibinfo {author} {\bibfnamefont
  {W.}~\bibnamefont {Weyrich}},\ }\href {\doibase 10.1515/zna-1993-1-241}
  {\bibfield  {journal} {\bibinfo  {journal} {Zeitschrift für Naturforschung
  A}\ }\textbf {\bibinfo {volume} {48}},\ \bibinfo {pages} {211} (\bibinfo
  {year} {1993})}\BibitemShut {NoStop}%
\bibitem [{\citenamefont {Schwarz}\ \emph {et~al.}(1994)\citenamefont
  {Schwarz}, \citenamefont {Langenbach},\ and\ \citenamefont
  {Birlenbach}}]{schwarzDensityMatricesPosition1994b}%
  \BibitemOpen
  \bibfield  {author} {\bibinfo {author} {\bibfnamefont {W.~H.~E.}\
  \bibnamefont {Schwarz}}, \bibinfo {author} {\bibfnamefont {A.}~\bibnamefont
  {Langenbach}}, \ and\ \bibinfo {author} {\bibfnamefont {L.}~\bibnamefont
  {Birlenbach}},\ }\href {\doibase 10.1007/BF01113293} {\bibfield  {journal}
  {\bibinfo  {journal} {Theoret. Chim. Acta}\ }\textbf {\bibinfo {volume}
  {88}},\ \bibinfo {pages} {437} (\bibinfo {year} {1994})}\BibitemShut
  {NoStop}%
\bibitem [{\citenamefont {Schmider}(1996)}]{schmiderLowMomentumElectrons1996}%
  \BibitemOpen
  \bibfield  {author} {\bibinfo {author} {\bibfnamefont {H.}~\bibnamefont
  {Schmider}},\ }\href {\doibase 10.1063/1.472233} {\bibfield  {journal}
  {\bibinfo  {journal} {J. Chem. Phys}\ }\textbf {\bibinfo {volume} {105}},\
  \bibinfo {pages} {3627} (\bibinfo {year} {1996})}\BibitemShut {NoStop}%
\bibitem [{\citenamefont {Gueddida}\ \emph
  {et~al.}(2018{\natexlab{a}})\citenamefont {Gueddida}, \citenamefont {Yan},\
  and\ \citenamefont {Gillet}}]{gueddidaDevelopmentJointRefinement2018}%
  \BibitemOpen
  \bibfield  {author} {\bibinfo {author} {\bibfnamefont {S.}~\bibnamefont
  {Gueddida}}, \bibinfo {author} {\bibfnamefont {Z.}~\bibnamefont {Yan}}, \
  and\ \bibinfo {author} {\bibfnamefont {J.-M.}\ \bibnamefont {Gillet}},\
  }\href {\doibase 10.1107/S2053273318000384} {\bibfield  {journal} {\bibinfo
  {journal} {Acta Cryst. A}\ }\textbf {\bibinfo {volume} {74}},\ \bibinfo
  {pages} {131} (\bibinfo {year} {2018}{\natexlab{a}})}\BibitemShut {NoStop}%
\bibitem [{\citenamefont {Gueddida}\ \emph
  {et~al.}(2018{\natexlab{b}})\citenamefont {Gueddida}, \citenamefont {Yan},
  \citenamefont {Kibalin}, \citenamefont {Voufack}, \citenamefont {Claiser},
  \citenamefont {Souhassou}, \citenamefont {Lecomte}, \citenamefont {Gillon},\
  and\ \citenamefont {Gillet}}]{gueddidaJointRefinementModel2018}%
  \BibitemOpen
  \bibfield  {author} {\bibinfo {author} {\bibfnamefont {S.}~\bibnamefont
  {Gueddida}}, \bibinfo {author} {\bibfnamefont {Z.}~\bibnamefont {Yan}},
  \bibinfo {author} {\bibfnamefont {I.}~\bibnamefont {Kibalin}}, \bibinfo
  {author} {\bibfnamefont {A.~B.}\ \bibnamefont {Voufack}}, \bibinfo {author}
  {\bibfnamefont {N.}~\bibnamefont {Claiser}}, \bibinfo {author} {\bibfnamefont
  {M.}~\bibnamefont {Souhassou}}, \bibinfo {author} {\bibfnamefont
  {C.}~\bibnamefont {Lecomte}}, \bibinfo {author} {\bibfnamefont
  {B.}~\bibnamefont {Gillon}}, \ and\ \bibinfo {author} {\bibfnamefont {J.-M.}\
  \bibnamefont {Gillet}},\ }\href {\doibase 10.1063/1.5022770} {\bibfield
  {journal} {\bibinfo  {journal} {J. Chem. Phys}\ }\textbf {\bibinfo {volume}
  {148}},\ \bibinfo {pages} {164106} (\bibinfo {year}
  {2018}{\natexlab{b}})}\BibitemShut {NoStop}%
\bibitem [{\citenamefont {De~Bruyne}\ and\ \citenamefont
  {Gillet}(2020)}]{debruyneInferringOneelectronReduced2020}%
  \BibitemOpen
  \bibfield  {author} {\bibinfo {author} {\bibfnamefont {B.}~\bibnamefont
  {De~Bruyne}}\ and\ \bibinfo {author} {\bibfnamefont {J.-M.}\ \bibnamefont
  {Gillet}},\ }\href {\doibase 10.1107/S2053273319015870} {\bibfield  {journal}
  {\bibinfo  {journal} {Acta Cryst. A}\ }\textbf {\bibinfo {volume} {76}},\
  \bibinfo {pages} {1} (\bibinfo {year} {2020})}\BibitemShut {NoStop}%
\bibitem [{\citenamefont {Launay}\ and\ \citenamefont
  {Gillet}(2021)}]{launayNRepresentableOneelectronReduced2021}%
  \BibitemOpen
  \bibfield  {author} {\bibinfo {author} {\bibfnamefont {Y.}~\bibnamefont
  {Launay}}\ and\ \bibinfo {author} {\bibfnamefont {J.-M.}\ \bibnamefont
  {Gillet}},\ }\href {\doibase 10.1107/S2052520621007228} {\bibfield  {journal}
  {\bibinfo  {journal} {Acta Cryst. B}\ }\textbf {\bibinfo {volume} {77}},\
  \bibinfo {pages} {683} (\bibinfo {year} {2021})}\BibitemShut {NoStop}%
\bibitem [{\citenamefont
  {Löwdin}(1955)}]{lowdinQuantumTheoryManyParticle1955a}%
  \BibitemOpen
  \bibfield  {author} {\bibinfo {author} {\bibfnamefont {P.-O.}\ \bibnamefont
  {Löwdin}},\ }\href {\doibase 10.1103/PhysRev.97.1474} {\bibfield  {journal}
  {\bibinfo  {journal} {Phys. Rev.}\ }\textbf {\bibinfo {volume} {97}},\
  \bibinfo {pages} {1474} (\bibinfo {year} {1955})}\BibitemShut {NoStop}%
\bibitem [{\citenamefont {Weyrich}(1996)}]{Weyrich1996}%
  \BibitemOpen
  \bibfield  {author} {\bibinfo {author} {\bibfnamefont {W.}~\bibnamefont
  {Weyrich}},\ }\enquote {\bibinfo {title} {One-electron density matrices and
  related observables},}\ in\ \href {\doibase 10.1007/978-3-642-61478-1_14}
  {\emph {\bibinfo {booktitle} {Quantum-Mechanical Ab-initio Calculation of the
  Properties of Crystalline Materials}}}\ (\bibinfo  {publisher} {Springer
  Berlin Heidelberg},\ \bibinfo {address} {Berlin, Heidelberg},\ \bibinfo
  {year} {1996})\ pp.\ \bibinfo {pages} {245--272}\BibitemShut {NoStop}%
\bibitem [{\citenamefont {Gatti}\ and\ \citenamefont
  {Macchi}(2012)}]{gattiModernChargeDensityAnalysis2012a}%
  \BibitemOpen
  \bibinfo {editor} {\bibfnamefont {C.}~\bibnamefont {Gatti}}\ and\ \bibinfo
  {editor} {\bibfnamefont {P.}~\bibnamefont {Macchi}},\ eds.,\ \href {\doibase
  10.1007/978-90-481-3836-4} {\emph {\bibinfo {title} {Modern
  {Charge}-{Density} {Analysis}}}}\ (\bibinfo  {publisher} {Springer
  Netherlands},\ \bibinfo {address} {Dordrecht},\ \bibinfo {year}
  {2012})\BibitemShut {NoStop}%
\bibitem [{\citenamefont {Phillips}\ and\ \citenamefont
  {Weiss}(1968)}]{phillipsXRayDeterminationElectron1968}%
  \BibitemOpen
  \bibfield  {author} {\bibinfo {author} {\bibfnamefont {W.~C.}\ \bibnamefont
  {Phillips}}\ and\ \bibinfo {author} {\bibfnamefont {R.~J.}\ \bibnamefont
  {Weiss}},\ }\href {\doibase 10.1103/PhysRev.171.790} {\bibfield  {journal}
  {\bibinfo  {journal} {Phys. Rev.}\ }\textbf {\bibinfo {volume} {171}},\
  \bibinfo {pages} {790} (\bibinfo {year} {1968})}\BibitemShut {NoStop}%
\bibitem [{\citenamefont {{Wolfram Research}}(2023)}]{Mathematica}%
  \BibitemOpen
  \bibfield  {author} {\bibinfo {author} {\bibnamefont {{Wolfram Research}}},\
  }\href@noop {} {\enquote {\bibinfo {title} {Mathematica, {V}ersion 13.3},}\ }
  (\bibinfo {year} {2023}),\ \bibinfo {note} {champaign, IL, 2023}\BibitemShut
  {NoStop}%
\bibitem [{\citenamefont {Boyd}\ and\ \citenamefont
  {Vandenberghe}(2004)}]{boyd2004convex}%
  \BibitemOpen
  \bibfield  {author} {\bibinfo {author} {\bibfnamefont {S.~P.}\ \bibnamefont
  {Boyd}}\ and\ \bibinfo {author} {\bibfnamefont {L.}~\bibnamefont
  {Vandenberghe}},\ }\href@noop {} {\emph {\bibinfo {title} {Convex
  optimization}}}\ (\bibinfo  {publisher} {Cambridge university press},\
  \bibinfo {year} {2004})\BibitemShut {NoStop}%
\bibitem [{\citenamefont {Chakraborty}\ and\ \citenamefont
  {Mazziotti}(2015)}]{Chakraborty2015Nrep}%
  \BibitemOpen
  \bibfield  {author} {\bibinfo {author} {\bibfnamefont {R.}~\bibnamefont
  {Chakraborty}}\ and\ \bibinfo {author} {\bibfnamefont {D.~A.}\ \bibnamefont
  {Mazziotti}},\ }\href {\doibase https://doi.org/10.1002/qua.24934} {\bibfield
   {journal} {\bibinfo  {journal} {Int. J. Quantum Chem.}\ }\textbf {\bibinfo
  {volume} {115}},\ \bibinfo {pages} {1305} (\bibinfo {year}
  {2015})}\BibitemShut {NoStop}%
\bibitem [{\citenamefont {Diamond}\ and\ \citenamefont
  {Boyd}(2016)}]{diamond2016cvxpy}%
  \BibitemOpen
  \bibfield  {author} {\bibinfo {author} {\bibfnamefont {S.}~\bibnamefont
  {Diamond}}\ and\ \bibinfo {author} {\bibfnamefont {S.}~\bibnamefont {Boyd}},\
  }\href@noop {} {\bibfield  {journal} {\bibinfo  {journal} {J. Mach. Learn.
  Res}\ }\textbf {\bibinfo {volume} {17}},\ \bibinfo {pages} {1} (\bibinfo
  {year} {2016})}\BibitemShut {NoStop}%
\bibitem [{\citenamefont {Sternemann}\ \emph {et~al.}(2000)\citenamefont
  {Sternemann}, \citenamefont {Döring}, \citenamefont {Wittkop}, \citenamefont
  {Schülke}, \citenamefont {Shukla}, \citenamefont {Buslaps},\ and\
  \citenamefont {Suortti}}]{sternemannInfluenceLatticeDynamics2000}%
  \BibitemOpen
  \bibfield  {author} {\bibinfo {author} {\bibfnamefont {C.}~\bibnamefont
  {Sternemann}}, \bibinfo {author} {\bibfnamefont {G.}~\bibnamefont {Döring}},
  \bibinfo {author} {\bibfnamefont {C.}~\bibnamefont {Wittkop}}, \bibinfo
  {author} {\bibfnamefont {W.}~\bibnamefont {Schülke}}, \bibinfo {author}
  {\bibfnamefont {A.}~\bibnamefont {Shukla}}, \bibinfo {author} {\bibfnamefont
  {T.}~\bibnamefont {Buslaps}}, \ and\ \bibinfo {author} {\bibfnamefont
  {P.}~\bibnamefont {Suortti}},\ }\href {\doibase
  10.1016/S0022-3697(99)00321-2} {\bibfield  {journal} {\bibinfo  {journal} {J.
  Phys. Chem. Solids.}\ }\textbf {\bibinfo {volume} {61}},\ \bibinfo {pages}
  {379} (\bibinfo {year} {2000})}\BibitemShut {NoStop}%
\bibitem [{\citenamefont {Dugdale}\ and\ \citenamefont
  {Jarlborg}(1998)}]{dugdaleThermalDisorderCorrelation1998}%
  \BibitemOpen
  \bibfield  {author} {\bibinfo {author} {\bibfnamefont {S.}~\bibnamefont
  {Dugdale}}\ and\ \bibinfo {author} {\bibfnamefont {T.}~\bibnamefont
  {Jarlborg}},\ }\href {\doibase https://doi.org/10.1016/S0038-1098(97)10112-0}
  {\bibfield  {journal} {\bibinfo  {journal} {Solid State Commun.}\ }\textbf
  {\bibinfo {volume} {105}},\ \bibinfo {pages} {283} (\bibinfo {year}
  {1998})}\BibitemShut {NoStop}%
\bibitem [{\citenamefont {Matsuda}\ \emph {et~al.}(2020)\citenamefont
  {Matsuda}, \citenamefont {Kimura}, \citenamefont {Hagiya}, \citenamefont
  {Kajihara}, \citenamefont {Inui}, \citenamefont {Hiraoka}, \citenamefont
  {Tamura},\ and\ \citenamefont {Sakurai}}]{matsudaRayComptonScattering2020}%
  \BibitemOpen
  \bibfield  {author} {\bibinfo {author} {\bibfnamefont {K.}~\bibnamefont
  {Matsuda}}, \bibinfo {author} {\bibfnamefont {K.}~\bibnamefont {Kimura}},
  \bibinfo {author} {\bibfnamefont {T.}~\bibnamefont {Hagiya}}, \bibinfo
  {author} {\bibfnamefont {Y.}~\bibnamefont {Kajihara}}, \bibinfo {author}
  {\bibfnamefont {M.}~\bibnamefont {Inui}}, \bibinfo {author} {\bibfnamefont
  {N.}~\bibnamefont {Hiraoka}}, \bibinfo {author} {\bibfnamefont
  {K.}~\bibnamefont {Tamura}}, \ and\ \bibinfo {author} {\bibfnamefont
  {Y.}~\bibnamefont {Sakurai}},\ }\href {\doibase 10.1002/pssb.202000187}
  {\bibfield  {journal} {\bibinfo  {journal} {Phys. Status Solidi B}\ }\textbf
  {\bibinfo {volume} {257}},\ \bibinfo {pages} {2000187} (\bibinfo {year}
  {2020})}\BibitemShut {NoStop}%
\bibitem [{\citenamefont {Stevens}\ \emph {et~al.}(1977)\citenamefont
  {Stevens}, \citenamefont {Rys},\ and\ \citenamefont
  {Coppens}}]{stevensCalculationDynamicElectron1977}%
  \BibitemOpen
  \bibfield  {author} {\bibinfo {author} {\bibfnamefont {E.~D.}\ \bibnamefont
  {Stevens}}, \bibinfo {author} {\bibfnamefont {J.}~\bibnamefont {Rys}}, \ and\
  \bibinfo {author} {\bibfnamefont {P.}~\bibnamefont {Coppens}},\ }\href
  {\doibase 10.1107/S0567739477000801} {\bibfield  {journal} {\bibinfo
  {journal} {Acta Cryst. A}\ }\textbf {\bibinfo {volume} {33}},\ \bibinfo
  {pages} {333} (\bibinfo {year} {1977})}\BibitemShut {NoStop}%
\bibitem [{\citenamefont {Erba}\ \emph {et~al.}(2013)\citenamefont {Erba},
  \citenamefont {Ferrabone}, \citenamefont {Orlando},\ and\ \citenamefont
  {Dovesi}}]{erbaAccurateDynamicalStructure2013a}%
  \BibitemOpen
  \bibfield  {author} {\bibinfo {author} {\bibfnamefont {A.}~\bibnamefont
  {Erba}}, \bibinfo {author} {\bibfnamefont {M.}~\bibnamefont {Ferrabone}},
  \bibinfo {author} {\bibfnamefont {R.}~\bibnamefont {Orlando}}, \ and\
  \bibinfo {author} {\bibfnamefont {R.}~\bibnamefont {Dovesi}},\ }\href
  {\doibase 10.1002/jcc.23138} {\bibfield  {journal} {\bibinfo  {journal} {J.
  Comput. Chem.}\ }\textbf {\bibinfo {volume} {34}},\ \bibinfo {pages} {346}
  (\bibinfo {year} {2013})}\BibitemShut {NoStop}%
\bibitem [{\citenamefont {Cassidy}\ \emph {et~al.}(1979)\citenamefont
  {Cassidy}, \citenamefont {Halbout}, \citenamefont {Donaldson},\ and\
  \citenamefont {Tang}}]{cassidyNonlinearOpticalProperties1979}%
  \BibitemOpen
  \bibfield  {author} {\bibinfo {author} {\bibfnamefont {C.}~\bibnamefont
  {Cassidy}}, \bibinfo {author} {\bibfnamefont {J.}~\bibnamefont {Halbout}},
  \bibinfo {author} {\bibfnamefont {W.}~\bibnamefont {Donaldson}}, \ and\
  \bibinfo {author} {\bibfnamefont {C.}~\bibnamefont {Tang}},\ }\href {\doibase
  10.1016/0030-4018(79)90027-0} {\bibfield  {journal} {\bibinfo  {journal}
  {Opt. Commun.}\ }\textbf {\bibinfo {volume} {29}},\ \bibinfo {pages} {243}
  (\bibinfo {year} {1979})}\BibitemShut {NoStop}%
\bibitem [{\citenamefont {West}\ \emph {et~al.}(2015)\citenamefont {West},
  \citenamefont {Schmidt}, \citenamefont {Gordon},\ and\ \citenamefont
  {Ruedenberg}}]{westComprehensiveAnalysisTerms2015}%
  \BibitemOpen
  \bibfield  {author} {\bibinfo {author} {\bibfnamefont {A.~C.}\ \bibnamefont
  {West}}, \bibinfo {author} {\bibfnamefont {M.~W.}\ \bibnamefont {Schmidt}},
  \bibinfo {author} {\bibfnamefont {M.~S.}\ \bibnamefont {Gordon}}, \ and\
  \bibinfo {author} {\bibfnamefont {K.}~\bibnamefont {Ruedenberg}},\ }\href
  {\doibase 10.1021/acs.jpca.5b03400} {\bibfield  {journal} {\bibinfo
  {journal} {J. Phys. Chem. A}\ }\textbf {\bibinfo {volume} {119}},\ \bibinfo
  {pages} {10368} (\bibinfo {year} {2015})}\BibitemShut {NoStop}%
\bibitem [{\citenamefont {Zavodnik}\ \emph {et~al.}(1999)\citenamefont
  {Zavodnik}, \citenamefont {Stash}, \citenamefont {Tsirelson}, \citenamefont
  {De~Vries},\ and\ \citenamefont {Feil}}]{zavodnikElectronDensityStudy1999}%
  \BibitemOpen
  \bibfield  {author} {\bibinfo {author} {\bibfnamefont {V.}~\bibnamefont
  {Zavodnik}}, \bibinfo {author} {\bibfnamefont {A.}~\bibnamefont {Stash}},
  \bibinfo {author} {\bibfnamefont {V.}~\bibnamefont {Tsirelson}}, \bibinfo
  {author} {\bibfnamefont {R.}~\bibnamefont {De~Vries}}, \ and\ \bibinfo
  {author} {\bibfnamefont {D.}~\bibnamefont {Feil}},\ }\href {\doibase
  10.1107/S0108768198005746} {\bibfield  {journal} {\bibinfo  {journal} {Acta
  Cryst. B}\ }\textbf {\bibinfo {volume} {55}},\ \bibinfo {pages} {45}
  (\bibinfo {year} {1999})}\BibitemShut {NoStop}%
\bibitem [{\citenamefont {Birkedal}\ \emph {et~al.}(2004)\citenamefont
  {Birkedal}, \citenamefont {Madsen}, \citenamefont {Mathiesen}, \citenamefont
  {Knudsen}, \citenamefont {Weber}, \citenamefont {Pattison},\ and\
  \citenamefont {Schwarzenbach}}]{birkedalChargeDensityUrea2004}%
  \BibitemOpen
  \bibfield  {author} {\bibinfo {author} {\bibfnamefont {H.}~\bibnamefont
  {Birkedal}}, \bibinfo {author} {\bibfnamefont {D.}~\bibnamefont {Madsen}},
  \bibinfo {author} {\bibfnamefont {R.~H.}\ \bibnamefont {Mathiesen}}, \bibinfo
  {author} {\bibfnamefont {K.}~\bibnamefont {Knudsen}}, \bibinfo {author}
  {\bibfnamefont {H.-P.}\ \bibnamefont {Weber}}, \bibinfo {author}
  {\bibfnamefont {P.}~\bibnamefont {Pattison}}, \ and\ \bibinfo {author}
  {\bibfnamefont {D.}~\bibnamefont {Schwarzenbach}},\ }\href {\doibase
  10.1107/S0108767304015120} {\bibfield  {journal} {\bibinfo  {journal} {Acta
  Cryst. A}\ }\textbf {\bibinfo {volume} {60}},\ \bibinfo {pages} {371}
  (\bibinfo {year} {2004})}\BibitemShut {NoStop}%
\bibitem [{\citenamefont {Shukla}\ \emph {et~al.}(2001)\citenamefont {Shukla},
  \citenamefont {Isaacs}, \citenamefont {Hamann},\ and\ \citenamefont
  {Platzman}}]{shuklaHydrogenBondingUrea2001}%
  \BibitemOpen
  \bibfield  {author} {\bibinfo {author} {\bibfnamefont {A.}~\bibnamefont
  {Shukla}}, \bibinfo {author} {\bibfnamefont {E.~D.}\ \bibnamefont {Isaacs}},
  \bibinfo {author} {\bibfnamefont {D.~R.}\ \bibnamefont {Hamann}}, \ and\
  \bibinfo {author} {\bibfnamefont {P.~M.}\ \bibnamefont {Platzman}},\ }\href
  {\doibase 10.1103/PhysRevB.64.052101} {\bibfield  {journal} {\bibinfo
  {journal} {Phys. Rev. B}\ }\textbf {\bibinfo {volume} {64}},\ \bibinfo
  {pages} {052101} (\bibinfo {year} {2001})}\BibitemShut {NoStop}%
\bibitem [{\citenamefont
  {Becke}(1993)}]{beckeDensityFunctionalThermochemistry1993}%
  \BibitemOpen
  \bibfield  {author} {\bibinfo {author} {\bibfnamefont {A.~D.}\ \bibnamefont
  {Becke}},\ }\href {\doibase 10.1063/1.464913} {\bibfield  {journal} {\bibinfo
   {journal} {J. Chem. Phys}\ }\textbf {\bibinfo {volume} {98}},\ \bibinfo
  {pages} {5648} (\bibinfo {year} {1993})}\BibitemShut {NoStop}%
\bibitem [{\citenamefont {Peintinger}\ \emph {et~al.}(2013)\citenamefont
  {Peintinger}, \citenamefont {Oliveira},\ and\ \citenamefont
  {Bredow}}]{peintingerConsistentGaussianBasis2013}%
  \BibitemOpen
  \bibfield  {author} {\bibinfo {author} {\bibfnamefont {M.~F.}\ \bibnamefont
  {Peintinger}}, \bibinfo {author} {\bibfnamefont {D.~V.}\ \bibnamefont
  {Oliveira}}, \ and\ \bibinfo {author} {\bibfnamefont {T.}~\bibnamefont
  {Bredow}},\ }\href {\doibase 10.1002/jcc.23153} {\bibfield  {journal}
  {\bibinfo  {journal} {J. Comput. Chem.}\ }\textbf {\bibinfo {volume} {34}},\
  \bibinfo {pages} {451} (\bibinfo {year} {2013})}\BibitemShut {NoStop}%
\bibitem [{\citenamefont {Dovesi}\ \emph {et~al.}(2014)\citenamefont {Dovesi},
  \citenamefont {Orlando}, \citenamefont {Erba}, \citenamefont
  {Zicovich-Wilson}, \citenamefont {Civalleri}, \citenamefont {Casassa},
  \citenamefont {Maschio}, \citenamefont {Ferrabone}, \citenamefont
  {De~La~Pierre}, \citenamefont {D'Arco}, \citenamefont {Noël}, \citenamefont
  {Causà}, \citenamefont {Rérat},\ and\ \citenamefont
  {Kirtman}}]{Dovesi2014Crystal}%
  \BibitemOpen
  \bibfield  {author} {\bibinfo {author} {\bibfnamefont {R.}~\bibnamefont
  {Dovesi}}, \bibinfo {author} {\bibfnamefont {R.}~\bibnamefont {Orlando}},
  \bibinfo {author} {\bibfnamefont {A.}~\bibnamefont {Erba}}, \bibinfo {author}
  {\bibfnamefont {C.~M.}\ \bibnamefont {Zicovich-Wilson}}, \bibinfo {author}
  {\bibfnamefont {B.}~\bibnamefont {Civalleri}}, \bibinfo {author}
  {\bibfnamefont {S.}~\bibnamefont {Casassa}}, \bibinfo {author} {\bibfnamefont
  {L.}~\bibnamefont {Maschio}}, \bibinfo {author} {\bibfnamefont
  {M.}~\bibnamefont {Ferrabone}}, \bibinfo {author} {\bibfnamefont
  {M.}~\bibnamefont {De~La~Pierre}}, \bibinfo {author} {\bibfnamefont
  {P.}~\bibnamefont {D'Arco}}, \bibinfo {author} {\bibfnamefont
  {Y.}~\bibnamefont {Noël}}, \bibinfo {author} {\bibfnamefont
  {M.}~\bibnamefont {Causà}}, \bibinfo {author} {\bibfnamefont
  {M.}~\bibnamefont {Rérat}}, \ and\ \bibinfo {author} {\bibfnamefont
  {B.}~\bibnamefont {Kirtman}},\ }\href {\doibase
  https://doi.org/10.1002/qua.24658} {\bibfield  {journal} {\bibinfo  {journal}
  {Int. J. Quantum Chem.}\ }\textbf {\bibinfo {volume} {114}},\ \bibinfo
  {pages} {1287} (\bibinfo {year} {2014})}\BibitemShut {NoStop}%
\bibitem [{\citenamefont {Worsham}\ \emph {et~al.}(1957)\citenamefont
  {Worsham}, \citenamefont {Levy},\ and\ \citenamefont
  {Peterson}}]{Worsham1986}%
  \BibitemOpen
  \bibfield  {author} {\bibinfo {author} {\bibfnamefont {J.~E.}\ \bibnamefont
  {Worsham}}, \bibinfo {author} {\bibfnamefont {H.~A.}\ \bibnamefont {Levy}}, \
  and\ \bibinfo {author} {\bibfnamefont {S.~W.}\ \bibnamefont {Peterson}},\
  }\href {\doibase 10.1107/S0365110X57000924} {\bibfield  {journal} {\bibinfo
  {journal} {Acta Crystallographica}\ }\textbf {\bibinfo {volume} {10}},\
  \bibinfo {pages} {319} (\bibinfo {year} {1957})}\BibitemShut {NoStop}%
\bibitem [{\citenamefont {Hansen}\ and\ \citenamefont
  {Coppens}(1978)}]{hansenTestingAsphericalAtom1978}%
  \BibitemOpen
  \bibfield  {author} {\bibinfo {author} {\bibfnamefont {N.~K.}\ \bibnamefont
  {Hansen}}\ and\ \bibinfo {author} {\bibfnamefont {P.}~\bibnamefont
  {Coppens}},\ }\href {\doibase 10.1107/S0567739478001886} {\bibfield
  {journal} {\bibinfo  {journal} {Acta Cryst. A}\ }\textbf {\bibinfo {volume}
  {34}},\ \bibinfo {pages} {909} (\bibinfo {year} {1978})}\BibitemShut
  {NoStop}%
\end{thebibliography}%

\onecolumngrid
\newpage
\section*{Supplementary Information}

\setcounter{figure}{0}
\renewcommand{\thefigure}{S\arabic{figure}}

\setcounter{table}{0}
\renewcommand{\thetable}{S\arabic{table}}

\setcounter{page}{1}
\subsection{Fitting score}

Here, we list the unweighted fitting scores $\chi^2$ for different data sets and reconstructions. We note $\tilde{\chi}^2_{\text{total}}$ the value of the objective function in the optimization procedure and $\chi^2_{\text{total}}$ the fitting score computed on the entire range of 0 K, noise-free artificial-data. 

Firstly, it can be observed that while the optimal objective function values for the 0 K and 50 K scenarios are similar, the $\chi^2_{\text{SF}}$ is substantially larger than the $\tilde{\chi}^2_{\text{SF}}$ in the 50 K case, suggesting that the thermal motion is not perfectly accounted for by \eqref{DWSFmatrix}. Secondly, in both the 0 K and 50 K cases, restrictions lead to an increase in the optimal fitting score $\tilde{\chi}^2_{\text{total}}$ (for noised 0 K and 50 K data), but a decrease of the ${\chi}^2_{\text{total}}$ (noise-free 0 K). Such an observation quantitatively confirms the assumption that restrictions act as regularization, thereby improving reconstruction results.

\begin{table}[!htp]
    \centering
    \begin{tabular}{|c||c|c|c||c|c|c|}
         \hline
         Scenario & $\tilde{\chi}^2_{\text{SF}}$ & $\tilde{\chi}^2_{\text{DCP}}$ & $\tilde{\chi}_{\text{total}}^2 $
         & $\chi^2_{\text{SF}}$ & $\chi^2_{\text{DCP}}$ & $\chi_{\text{total}}^2$
         \\
         \hline
         0 K 1\% noise & 2.683 & 2.084 & 4.766 & 1.798 & 0.871 & 2.669
         \\
         0 K 1\% noise sym. & 3.090 & 2.272 & 5.362 & 1.336 & 0.734 & 2.070
         \\
         0 K 1\% noise sym. \& core & 3.190 & 2.271 & 5.362 & 1.234 & 0.741 & 1.975
         \\
         50 K 1\% noise & 2.597 & 2.258 & 4.855 & 22.065 & 0.896 & 22.961
         \\
         50 K 1\% noise sym. & 2.964 & 2.662 & 5.626 & 17.375 & 0.724 & 18.099
         \\
         50 K 1\% noise sym. \& core & 3.144 & 2.741 & 5.885 & 14.176 & 0.757 & 14.933 
         \\
         \hline
    \end{tabular}
    \caption{$\tilde{\chi}^2_{\text{SF}} = \sum_\mathbf{q} (\tilde{F}(\mathbf{q}) - \text{Tr} ( \mathbf{P} \mathbf{F}_\mathbf{q}) )^2 $ for $\vert \mathbf{q} \vert < 0.7$ and $\tilde{F}$ being the 1\% noisy artificial-data. Similarly, $\tilde{\chi}^2_{\text{DCP}}$ is the objective function for noisy data. As a reminder,  $\tilde{\chi}^2_{\text{total}} = \tilde{\chi}^2_{\text{SF}} + \tilde{\chi}^2_{\text{DCP}}$ is the objective function being optimized, and the noised SF artificial-data is at respectively 0 K and 50 K for two different scenarios. $\chi^2_{\text{total}}, \chi^2_{\text{SF}}, \chi^2_{\text{DCP}}$ refer to the fitting score with respect to the noise-free artificial-data using the entire Ewald sphere up to 1.1 \invangs (always at 0 K).}
    \label{tab:my_label}
\end{table}

\subsection{Mean-field energies}

We list the energies evaluated for different reconstructions. Note that the potential energies, hence the total energies, are calculated with a Hartree-Fock Hamiltonian operator, which allows us to calculate an approximate energy with 1-RDMs. For reference, the energies of the DFT 1-RDM with the pob-DZVP basis set are $T_{\text{DFT}}  = 223.315, ~ E_{\text{DFT}} = -225.005$. However, one should keep in mind that these energies are evaluated with different methods and different basis functions, and are not meant to be compared directly.

\begin{table}[!htp]
    \centering
    \begin{tabular}{|c||c|c|c||c|c|}
         \hline
         Scenario & $T$ & $V_{\text{HF}}$ & $E_{\text{HF}}$
         & Virial ratio
         \\
         \hline
         0 K 1\% noise &  224.554 & -447.229 & -222.675 & 0.99581
         \\
         0 K 1\% noise sym. & 224.387 & -447.546 & -223.159 & 0.99726
         \\
         0 K 1\% noise sym. \& core & 224.909 & -448.237 & -223.327 & 0.99648
         \\
         50 K 1\% noise & 223.844 & -445.405 & -221.561 & 0.99490
         \\
         50 K 1\% noise sym. & 223.676 & -446.073 & -222.397 & 0.99714
         \\
         50 K 1\% noise sym. \& core & 224.884 & -447.832 & -222.949 & 0.99570
         \\
         \hline
    \end{tabular}
    \caption{The energies and Virial ratios evaluated for different reconstructed 1-RDMs. For potential energies $V_{\text{HF}}$, a mean-field Hartree-Fock energy operator is employed.}
    \label{tab:my_label}
\end{table}

\end{document}